%% file: TeX_File.tex
\documentclass[journal]{IEEEtran}
\makeatletter

\usepackage{blindtext}
\usepackage[T1]{fontenc}
\usepackage{textcomp}
\usepackage{amsmath}
\usepackage{amssymb}
\usepackage{bm}
\usepackage{mdwmath}
\usepackage{cases}
\usepackage[short, c2]{optidef}
\usepackage{graphicx}
\graphicspath{{PDF/}{Biography/PDF/}}
\usepackage[dvipsnames]{xcolor}
\definecolor{myblue}{rgb}{0,0.4980,1} 
\definecolor{myred}{rgb}{0.8706,0.1608,0.0627} 
\usepackage[caption=false,font=footnotesize,subrefformat=parens]{subfig}
\usepackage[nosort,noadjust]{cite}
\usepackage{url}
\usepackage{tabularray}
\UseTblrLibrary{counter}
\usepackage[linesnumbered,ruled]{algorithm2e}

\usepackage{hyperref}
\newcommand{\colorhypersetup}{\@ifpackageloaded{hyperref}{\hypersetup{%
	bookmarksopen=true,%
	bookmarksnumbered=true,%
	pdfpagemode={UseOutlines},
	pdfstartview={FitH},%
	colorlinks=true,%
	linkcolor={myred},%
	citecolor={orange}
}}{\empty}}
\newcommand{\blackhypersetup}{\@ifpackageloaded{hyperref}{\hypersetup{%
	bookmarksopen=true,%
	bookmarksnumbered=true,%
	pdfpagemode={UseOutlines},
	pdfstartview={FitH},%
	colorlinks=true,%
	allcolors={black}
}}{\empty}}
 \blackhypersetup

\usepackage{mfirstuc-english}
\MFUhyphentrue
\usepackage[version3,description,IEEEtran]{eacro}
\acsetup{%
    pages/display=all,%
    pages/seq/use=false,%
    pages/name=true,%
    pages/fill={\quad},%
    make-links=true%
}
\input{Abbreviation_Definitions}

\newtheorem{theorem}{\textbf{Theorem}}

\newcommand{\upperroman}[1]{\uppercase\expandafter{\romannumeral#1}}

\newcommand{\myvec}[1]{\bm{\mathrm{#1}}}

\newcommand{\myunit}[1]{%
	\ifmmode
		\mathrm{#1}
	\else
		$ \mathrm{#1} $
	\fi}
\newcommand{\murm}{%
	\ifmmode
		\text{\textmu}
	\else
		\textmu
	\fi}

\newcommand{\MYnewpage}{%
	\ifCLASSOPTIONonecolumn
		\ifCLASSOPTIONjournal
			\typeout{The onecolumn journal mode.}
			\newpage
		\fi
	\fi}

\newlength{\mysinglefigwidth}
\newlength{\mymultifigwidth}
\ifCLASSOPTIONonecolumn
	\AtBeginDocument{\setlength{\mysinglefigwidth}{0.7\linewidth}}
\else
	\AtBeginDocument{\setlength{\mysinglefigwidth}{\linewidth}}
\fi
\ifCLASSOPTIONonecolumn
	\AtBeginDocument{\setlength{\mymultifigwidth}{0.5\linewidth}}
\else
	\AtBeginDocument{\setlength{\mymultifigwidth}{\linewidth}}
\fi

\makeatother
\begin{document}
\ifCLASSOPTIONonecolumn
    \typeout{The onecolumn mode.}
    \title{\LARGE Title}
    \author{Author~1,~\IEEEmembership{Member,~IEEE}, and~Author~2,~\IEEEmembership{Fellow,~IEEE}
        \thanks{Manuscript received XXX, 2019; revised XXX.}
        \thanks{Author~1 is with the Intelligent Computing and Communications ($ \text{IC}^\text{2} $) Lab, Wireless Signal Processing and Networks (WSPN) Lab, Key Laboratory of Universal Wireless Communications, Ministry of Education, Beijing University of Posts and Telecommunications (BUPT), Beijing, 100876, China (E-mail: \textsf{XXX@bupt.edu.cn}).}
        \thanks{Author~2 is with the Intelligent Computing and Communications ($ \text{IC}^\text{2} $) Lab, Wireless Signal Processing and Networks (WSPN) Lab, Key Laboratory of Universal Wireless Communications, Ministry of Education, Beijing University of Posts and Telecommunications (BUPT), Beijing, 100876, China (E-mail: \textsf{XXX@bupt.edu.cn}).}
        \thanks{This work was supported by the National Natural Science Foundation of China (NSFC) under Grant No. XXX.}
    }
\else
    \typeout{The twocolumn mode.}
    \title{Digital Twin-Based Network Management for Better QoE in Multicast Short Video Streaming}
    \author{Xinyu~Huang,~\IEEEmembership{Student~Member,~IEEE}, Shisheng~Hu,~\IEEEmembership{Student~Member,~IEEE}, Haojun~Yang,~\IEEEmembership{Member,~IEEE}, Xinghan~Wang,~\IEEEmembership{Student~Member,~IEEE}, Yingying~Pei,~\IEEEmembership{Student~Member,~IEEE}, and~Xuemin~(Sherman)~Shen,~\IEEEmembership{Fellow,~IEEE} 
        \thanks{Xinyu Huang, Shisheng Hu, Haojun Yang, Xinghan Wang, Yingying Pei, and Xuemin (Sherman) Shen are with the Department of Electrical and Computer Engineering, University of
	Waterloo, Waterloo, ON N2L 3G1, Canada (E-mail:\{x357huan, s97hu, h88yang, x243wang, y32pei, sshen\}@uwaterloo.ca).}
    }
\fi

\ifCLASSOPTIONonecolumn
	\typeout{The onecolumn mode.}
\else
	\typeout{The twocolumn mode.}
\fi

\maketitle

\ifCLASSOPTIONonecolumn
	\typeout{The onecolumn mode.}
	\vspace*{-50pt}
\else
	\typeout{The twocolumn mode.}
\fi
\begin{abstract}
Multicast short video streaming can enhance bandwidth utilization by enabling simultaneous video transmission to multiple users over shared wireless channels. The existing network management schemes mainly rely on the sequential buffering principle and general quality of experience (QoE) model, which may deteriorate QoE when users' swipe behaviors exhibit distinct spatiotemporal variation. In this paper, we propose a digital twin (DT)-based network management scheme to enhance QoE. Firstly, user status emulated by the DT is utilized to estimate the transmission capabilities and watching probability distributions of sub-multicast groups (SMGs) for an adaptive segment buffering. The SMGs' buffers are aligned to the unique virtual buffers managed by the DT for a fine-grained buffer update. Then, a multicast QoE model consisting of rebuffering time, video quality, and quality variation is developed, by considering the mutual influence of segment buffering among SMGs. Finally, a joint optimization problem of segment version selection and slot division is formulated to maximize QoE. To efficiently solve the problem, a data-model-driven algorithm is proposed by integrating a convex optimization method and a deep reinforcement learning algorithm. Simulation results based on the real-world dataset demonstrate that the proposed DT-based network management scheme outperforms benchmark schemes in terms of QoE improvement.
\end{abstract}

\ifCLASSOPTIONonecolumn
	\typeout{The onecolumn mode.}
	\vspace*{-10pt}
\else
	\typeout{The twocolumn mode.}
\fi

\IEEEpeerreviewmaketitle

\MYnewpage


\section{Introduction}
\label{sec:Introduction}

\IEEEPARstart{S}{hort} video platforms such as TikTok, Instagram Reels, and YouTube Shorts have experienced a dramatic surge in user scale, {with TikTok's monthly active users reaching 1.7 billion in 2023~\cite{report}}. Seamless video requests bring a huge traffic and computing burden to communication networks, {which cause frequent playback lags and video quality fluctuation, especially in areas with high user density, thereby deteriorating users' watching experience~\cite{jiang2021survey}.} Multicast transmission, {as an essential technology in wireless networks, can enable a single data stream to be disseminated to numerous users in a group simultaneously.} By multicasting short videos to the users with similar characteristics and geographical locations, the redundancy in video rendering, transcoding, and transmission can be effectively reduced, thereby alleviating the network traffic and computing burden~\cite{huang2023digital}.

For \ac{msvs} services, one of the key metrics to evaluate the performance is \ac{QoE}. It is typically a multi-dimensional metric used to quantify users' subjective and objective watching experience~\cite{8332093}. For instance, video quality (resolution, bitrate), latency, as well as rebuffering events are the most important factors that dominate the \ac{QoE} level~\cite{9525340}. To maintain users' \ac{QoE} at a high level, efficient network management is essential, {such as buffer management and resource scheduling}~\cite{8260530,9580665}. Specifically, due to the limited bandwidth and computing resources, {it needs to be determined which versions and quantities of segments (sequential components of a video sequence) should be transmitted to users' buffers}, and at what priority levels. Furthermore, due to asynchronous swipe behaviors in one \ac{mg}, one \ac{mg} can be further divided
into multiple \acp{smg}. The \ac{smg} with a leading video playback is defined as the leading \ac{smg}, otherwise, the lagging \ac{smg}. {The bandwidth and computing resources need to be flexibly and accurately assigned to each \ac{smg} to reduce service delay.} Nevertheless, existing network management schemes are mainly based on the sequential buffering principle and general \ac{QoE} model while neglecting the impact of users' swipe behaviors and the mutual influence of \acp{smg}' segment buffering. As a result, users may suffer from frequent playback lags and low video quality, thus leading to a low \ac{QoE}.

\Ac{dt} is a promising technique to optimize network management for better \ac{QoE}. \ac{dt} is defined as a full digital representation of a physical object, and real-time synchronization between the physical object and its corresponding digital replica~\cite{grieves2014digital}. {As an essential component embedded in the next-generation communication networks, {\ac{dt} is comprised of a data pool as well as several data processing and decision-making modules}, which can efficiently emulate users' behaviors and network conditions, abstract distilled features, and make network management decisions~\cite{huang2023, guo2023five}.} Owing to its powerful emulation, analysis, and decision-making capabilities, communication networks can more intelligently perceive \acp{smg}' behavior patterns, finely control \acp{smg}' buffer update, and provide customized network management strategies to enhance users' \ac{QoE}. In this work, the precise role of constructed \ac{dt} is to emulate users' future network conditions and swipe behaviors, abstract the watching probability distribution of segments, {and make tailored network management decisions based on the emulated user status and abstracted information.}

The motivation of the DT-based network management for \ac{msvs} includes three aspects. Firstly, due to users' stochastic swipe behaviors, users' viewing sequences are usually non-sequential. {The existing sequential buffering principle can cause the segments to be swiped to not being buffered in time}, resulting in playback lags. Moreover, users' buffer lengths are usually overestimated due to multicast segment buffering, {which can bring inaccurate rebuffering time estimation.} Therefore, it is essential to develop an efficient segment buffering scheme that adapts well to users' swipe behaviors. Secondly, the \ac{QoE} model is typically composed of multiple factors. Due to the impact of multicast transmission and segment buffering priority, different \ac{QoE} factors among \acp{smg} influence each other, collectively impacting the \ac{mg}'s \ac{QoE}. Therefore, it is paramount to establish a \ac{QoE} model specifically tailored for \ac{msvs}. Thirdly, {since network management is usually a multi-variable decision-making problem, such as the segment version selection and resource scheduling,} the interplay among the variables makes the optimization problem complex and difficult to solve. Therefore, {how to design an efficient algorithm to solve it for users' high \ac{QoE} is important.}

Designing an efficient DT-based network management scheme needs to address the following challenges: 1) incorporating the impact of users' swipe behaviors in multicast segment buffering; 2) establishing an accurate multicast QoE model; 3) designing an efficient algorithm to solve the complex multi-variable decision-making problem. Specifically, {users' swipe behaviors are stochastic and spatiotemporally varied, which are difficult to accurately predict in real time.} Therefore, how to conduct effective data abstraction to obtain the distilled swipe feature and utilize it to facilitate accurate multicast segment buffering is challenging.  Furthermore, since the lagging \ac{smg} can still receive segments from other leading \acp{smg} in its scheduling slot, this interactivity results in complex QoE estimation. Therefore, how to characterize the impact of multicast segment buffering among \acp{smg} on the \ac{QoE} model is challenging. Finally, since the network management problem is usually a mixed-integer non-convex problem, directly using model-based or data-driven algorithms can lead to a loss in the system performance. Therefore, how to ingeniously combine the advantages of model-based and data-driven algorithms to efficiently solve the formulated problem is challenging. 

In this paper, we propose a DT-based network management scheme, which can effectively enhance users' \ac{QoE}. The main contributions are summarized as follows:
\begin{itemize}
    \item 
    Firstly, we propose a novel \ac{dt}-assisted buffer management scheme to incorporate the impact of swipe behaviors. Specifically, {users' historical status, including locations, channel conditions, preferences and swipe timestamps, is stored in the \ac{dt} for status emulation.} The emulated status is used to abstract the \acp{smg}' transmission capabilities and the watching probability distribution of segments. Based on the abstracted information, \ac{dt} can make an adaptive segment buffering decision. Furthermore, {\ac{dt} is utilized to construct and manage virtual buffers for each \ac{smg} for a fine-grained buffer update.}
    \item 
    Secondly, we establish a multicast \ac{QoE} model to quantify the impact of multicast segment buffering among \acp{smg}. Specifically, the multicast \ac{QoE} model is built as a weighted sum of rebuffering time, video quality, and quality variation, where the weighting factors are the integration of buffering order and users' sensitivity degrees. {The rebuffering time estimation relies on the multicast transmission delay as well as the parallel transmission and transcoding process.} The video quality and quality variation depend on the relationship between the segment version and \ac{ssim}. Based on these elements, an accurate multicast \ac{QoE} model is established.
    \item 
    Thirdly, we formulate a joint optimization problem of segment version selection and slot division to maximize \ac{QoE}. Since the formulated problem is a mixed-integer nonlinear programming problem, it is hard to directly use a model-based or data-driven algorithm to solve it. Therefore, we propose a data-model-driven algorithm. Specifically, {a convex optimization method is embedded in a deep reinforcement learning (DRL) algorithm to decouple the joint optimization problem and reduce the action dimension, which can efficiently solve the formulated problem. The extensive simulation results on real-world short video streaming datasets show that the proposed DT-based network management scheme can effectively enhance QoE as compared with the state-of-the-art network management schemes.}
\end{itemize}

The remainder of this paper is organized as follows. Related works are introduced in Section~\ref{related}. The system model is first built in Section~\ref{syst}. Then, the \ac{dt}-assisted buffer management and the multicast \ac{QoE} model are presented in Sections~\ref{Buffer} and \ref{mqoe}, respectively. Next, the problem formulation and the proposed scheduling algorithm are shown in Section~\ref{problem}, followed by simulation results in Section~\ref{results}. Finally, Section~\ref{sec:Conclusion} concludes this paper.

\section{Related Work}\label{related}
{To facilitate the efficient \ac{msvs} within radio access networks (RANs)}, extensive works are devoted to optimizing network management performance from different directions, such as adaptive video bitrate, multi-connectivity management, and transmission and transcoding scheduling. Specifically, {Taha~\textit{et~al.} studied the impact of the characteristics of videos, wireless channel capabilities, and users' profiles on the \ac{mg}'s \ac{QoE}}, {and designed an efficient machine learning algorithm to adaptively adjust the video bitrate~\cite{taha2021qoe}.} {Lie \textit{et al.} integrated the \ac{svc} technology with the transmission delay constraint,} and exploited the hard deadline constrained prioritized data structure and user feedback to make an optimal adaptive encoding and scheduling strategy, which can enhance the average network throughput~\cite{7303965}. To achieve high-quality and cost-efficient multicast video services, Zhong~\textit{et~al.} proposed a novel buffer-nadir-based multicast mechanism, formulated the multicast-aware task offloading problem, and devised a joint optimization algorithm for data scheduling and task offloading, respectively~\cite{9741351}. Additionally, {Zuhra~\textit{et~al.} proposed the procedures of establishing multi-connectivity and a greedy approximation algorithm to solve the associated resource allocation problem}, which can effectively increase the number of served users~\cite{chaporkar2022leveraging}. Daher \textit{et al.} proposed a dynamic clustering algorithm based on the minimization of a submodular function that integrated the traffic in each cell and the \ac{mg}'s average \ac{sinr}, which can maintain an acceptable transmission failure probability and enhance the \ac{mg}'s average \ac{sinr}~\cite{9076859}. To improve the robustness of multicast transmission, {Zhang~\textit{et~al.} proposed a cooperative multicast framework, where users can recover videos with quality proportional to their channel conditions.} A joint power allocation and segment scheduling problem was formulated to minimize the overall distortion and solved by a provably convergent optimal algorithm~\cite{9756228}. Considering the characteristic of stochastic channels, Zhang~\textit{et~al.} further analyzed the video-layer recovery failure probability and estimated the \ac{mg}'s average \ac{QoE}. Based on the information, an optimal scheduling algorithm was developed based on the hidden monotonicity of the problem to maximize the \ac{mg}'s \ac{QoE}~\cite{9020157}. However, these schemes usually require real-time data collection and efficient data processing on user and network status information, which poses a high requirement for efficient network management.

As an essential virtualization technology, \ac{dt} was first introduced to monitor and mitigate anomalous events for flying vehicles~\cite{glaessgen2012digital}. {By introducing \ac{dt} into RANs, we can realize holistic network virtualization for efficient network management.} We refer readers to recent comprehensive surveys and tutorials on \ac{dt} to become familiar with this topic~\cite{holi, 10183802, 9939166}. There also exist some technical papers aiming at utilizing \ac{dt} to improve network management performance. Specifically, {Bellavista \textit{et al.} proposed an application-driven \ac{dt} networking middleware to simplify the interaction with heterogeneous distributed industrial devices and flexibly manage network resources}, which can effectively reduce communication overhead~\cite{bellavista2021application}. Qi \textit{et al.} leveraged the \ac{dt} as a centric controller to encourage edge devices to share their idle resources and get a reward from other devices with poor network performance, which can efficiently reduce the service delay~\cite{10234540}. Considering that the real networks are not static, {Dong~\textit{et~al.} constructed the \ac{dt} of the wireless networks to generate labeled training samples, where the network topology, channel and queueing models, and fundamental rules were adopted in the \ac{dt} to mirror the real networks}~\cite{8764584}. A \ac{dt}-assisted resource demand prediction scheme was proposed to enhance prediction accuracy for \ac{msvs}, including the \ac{dt} construction, the accurate and fast \ac{mg} construction, and the \ac{mg}'s swipe probability distribution abstraction~\cite{prework}. To further enhance users' \ac{QoE} in dynamic and heterogeneous environments, an intelligent resource allocation strategy with low communication overhead was proposed, {where \ac{dt} was utilized to monitor the current network operation status and enable intelligent decision-making~\cite{10234388}.} Furthermore, Jeremiah \textit{et al.} proposed to construct a central \ac{dt} to simulate dynamic and heterogeneous networks, which can enhance the efficiency of edge collaboration and real-time resource information availability~\cite{JEREMIAH2024243}. These pioneering works can effectively improve network management performance in terms of communication overhead, service delay, and \ac{QoE}.

Compared with related network management work, we have proposed a novel \ac{dt}-assisted buffer management scheme to incorporate the impact of swipe behaviors. Furthermore, we have established a multicast \ac{QoE} model to quantify the impact of multicast segment buffering among \acp{smg}. Finally, a data-model-driven algorithm has been developed to efficiently solve the formulated mixed-integer nonlinear programming problem for better \ac{QoE}.

\section{System Model}\label{syst}
\begin{figure}[!t]
    \centering
    \includegraphics[width=\mysinglefigwidth]{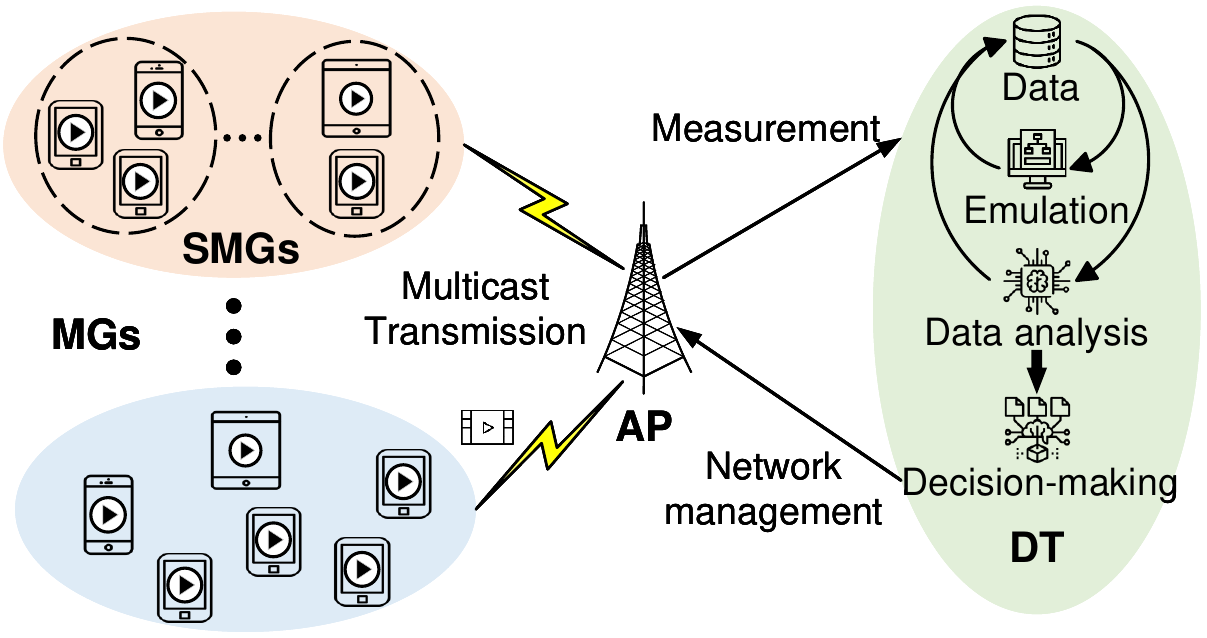}
    \caption{DT-assisted MSVS framework.}
    \label{scenario}
\end{figure}
As shown in Fig.~\ref{scenario}, we consider a DT-assisted MSVS scenario, which consists of an \ac{ap}, multiple \acp{mg}, and one \ac{dt}.
\begin{itemize}
    \item 
Access point: The \ac{ap} owns communication, computing, and caching capabilities. Based on users' requests, cached video sequences will be transcoded to appropriate bitrates and then transmitted to each \ac{mg}. In addition, it is responsible for collecting users' network-related and behavior-related information to update \ac{dt} data.  

\item
Multicast group: Each MG consists of multiple users using short video services. {The same video sequences will be transmitted from the AP to one MG over shared wireless channels.} {The \ac{mg}'s construction and update mainly depend on users' similarities in swipe behaviors, channel conditions, locations, and preferences}, as discussed in~\cite{huang2023digital}. Due to asynchronous swipe behaviors, users' devices in the same MG still have different playback stamps and buffer lengths. Therefore, one MG is further divided into multiple \acp{smg}, denoted by $\mathcal{G}=\{1,\cdots,G\}$. The number of total users in the system is denoted by $K$. The set of users in \ac{smg} $g$ is denoted by ${{\left\{ {{\mathcal{K}}_{g}} \right\}}_{g\in \mathcal{G}}}$.

\item
Digital twin: It consists of a database storing users' status information, a status emulation module, {as well as multiple data analysis and decision-making modules.} \ac{dt} data comes from two aspects, i.e., data measurement and emulation module. The former is used to obtain the label data to detect whether the emulated data is accurate enough and supplement \ac{dt} data when emulated data has a large deviation. {The latter uses machine learning-based algorithms to predict users' future status.} The data analysis module is responsible for abstracting distilled data features, such as swipe probability distribution, request density, user satisfaction, etc, {which can facilitate tailored network management in the decision-making module.}
\end{itemize}

\begin{figure}[!t]
    \centering
    \includegraphics[width=\mysinglefigwidth]{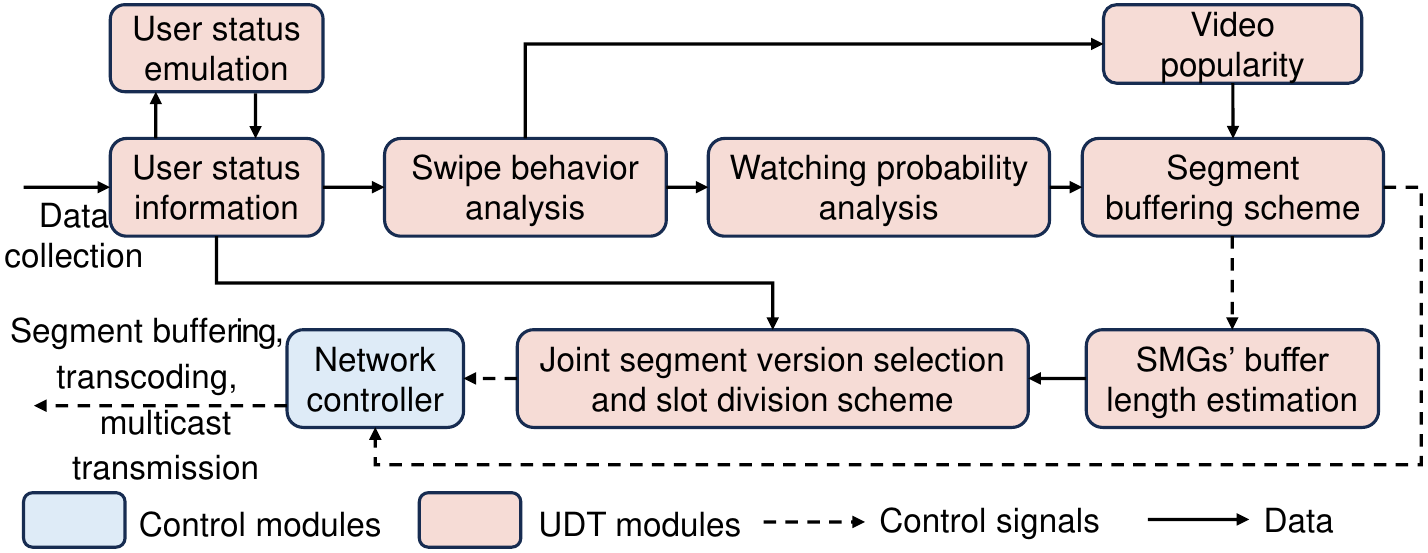}
    \caption{Proposed DT-assisted MSVS workflow.}
    \label{work}
\end{figure}

As shown in Fig.~\ref{work}, we present the proposed DT-assisted \ac{msvs} workflow. Specifically, the user status emulation module can generate user status information. When the generated data has a significant deviation from the actual user status data, new actual data is collected to correct user status information. The user status information is first utilized to abstract the user's swipe feature, such as swipe probability distribution, {which can update the video popularity and watching probability distribution of segments.} The updated information is then used to make the tailored segment buffering and \acp{smg}' buffer update decisions. Next, \acp{smg}' buffer lengths are estimated, which are integrated with user status information to make a joint segment version selection and slot division decision. Finally, all network management decisions are transferred to the network controller to guide the physical entities to implement.

\section{DT-Assisted Buffer Management Scheme}\label{Buffer}
In this section, we first propose an adaptive segment buffering scheme based on the DT-analyzed watching probability distribution, and then design a fine-grained buffer update scheme based on the DT-assisted virtual buffer management.

\subsection{DT Construction}
{DT consists of multiple modules, which are summarized as status emulation, data analysis, and decision-making, as shown in Fig.~\ref{work}.} 

Firstly, since network conditions and user behaviors are essential to reflect users' characteristics and requirements, we utilize the \ac{ap} to collect users' data from two aspects, i.e., networking-related data and behavior-related data. The networking-related data include users' channel conditions and locations, {which are utilized to estimate the transmission capabilities.} The behavior-related data consist of users' swipe timestamps and preferences. The long short-term memory (LSTM) method is utilized to mine data correlation, which can effectively emulate users' future networking-related data and behavior-related data. This operation can effectively reduce frequent data interaction costs between DT and users. By integrating these two kinds of data, DT can accurately emulate users' real-time status for network management.

Secondly, two kinds of data analysis modules are embedded in the DT, i.e., the swipe behavior analysis module and the watching probability analysis module. The former has been investigated in \cite{huang2023digital} for resource demand prediction. The latter is our focus in this paper, {which aims at assisting the buffer management.} Specifically, {due to the impact of sequential playback and swipe behaviors}, each segment's watching probability has a dependable relationship. From the aspects of the first and subsequent segments' swipe probabilities, the watching probability distribution of segments is derived in DT. Based on the derived information, DT can realize an accurate buffer management.

Thirdly, based on the emulated status and analyzed data, DT can realize tailored decision-makings to further enhance users' \ac{QoE}. Three kinds of decision-making modules are designed to make the segment buffering scheme, the \acp{smg}' buffer update scheme, and the joint segment version selection and slot division scheme, respectively. These modules are intricately coupled, where the output of one module seamlessly transitions into the input of the following module, collectively influencing QoE. {A data-model-driven algorithm is proposed to decouple the joint optimization problem and reduce the action dimension for efficient network management.}

\subsection{DT-Assisted Segment Buffering Scheme}
Due to users' diversified swipe behaviors, the segments to be watched do not follow a sequential order, which can cause different watching probabilities. To reduce rebuffering events, we need to analyze which segments should be prioritized for buffering. Therefore, we design a DT-assisted segment buffering scheme. The specific analysis is as follows.

\begin{figure}[!t]
    \centering
    \includegraphics[width=0.85\mysinglefigwidth]{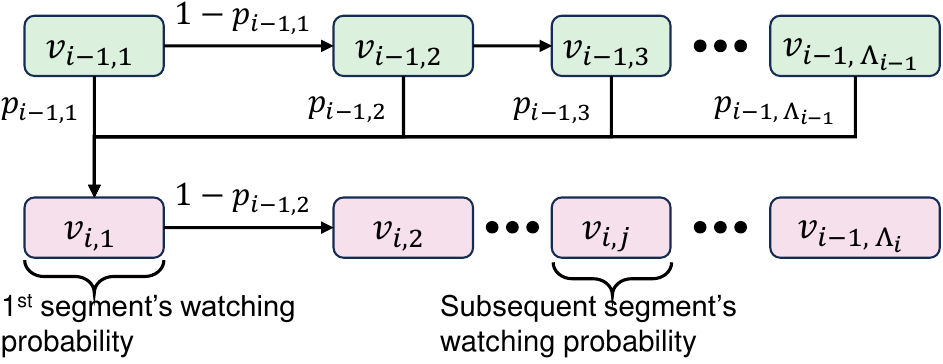}
    \caption{Watching probability analysis.}
    \label{wat-prob}
\end{figure}

As shown in Fig.~\ref{wat-prob}, \ac{dt} recommends a video list for the \ac{mg} in each large timescale, referred in \cite{huang2023digital}. Each video sequence consists of multiple segments with the same time length. The recommended video list is denoted by ${{\left\{ {{v}_{i,j}} \right\}}_{i\in \mathcal{N},j\in {\{1,\cdots,\Lambda_i \}}}}$, where $\mathcal{N}$ is the recommended video index list, and ${{\Lambda }_{i}}$ is the number of segments in video $i$. The watching probability distribution of segments has a dependable relationship~\cite{286427}. Let $w_{i,j}$ represent the watching probability of segment $v_{i,j}$. For any two successive segments $j$ and $j+1$ in video $i$, the watching probability of segment $j+1$ is derived as
\begin{equation}\label{watch}
w_{i,j+1} = w_{i,j}(1-p_{i,j}), \forall i\in \mathcal{N}, j\in \{1,\cdots,\Lambda_i - 1 \},
\end{equation}
where $p_{i,j}$ is the swipe probability of segment $v_{i,j}$. Based on Eq.~\eqref{watch}, we further analyze the watching probability distribution from the perspectives of $1^{\text{st}}$ and subsequent segments.

For the $1^{\text{st}}$ segment of video $i$, its watching probability is related to the swipe probability of all segments in previous video $i-1$. Therefore, the corresponding watching probability, ${{w}_{i,1}}$, is expressed as
\begin{equation}
    {{w}_{i,1}}={{w}_{i-1,1}}\left( {{p}_{i-1,1}}+\sum\limits_{j=1}^{{{\Lambda }_{i-1}}-1}{{{p}_{i-1,j+1}}\prod\limits_{k=1}^{j}{\left( 1-{{p}_{i-1,k}} \right)}} \right).
\end{equation}

For the subsequent segment $j$ of video $i$, its watching probability is only related to the swipe probability of previous segment $j-1$. Therefore, the corresponding watching probability, ${{w}_{i,j}}$, is given by
\begin{equation}
    {{w}_{i,j}}={{w}_{i,1}}\prod\limits_{k=1}^{j-1}{\left( 1-{{p}_{i,k}} \right)},\forall j\in \left\{ 2,\ldots ,{{\Lambda }_{i}} \right\}.
\end{equation}

Based on the watching probability distribution, we can determine the segment buffering order from high to low. {Furthermore, we analyze how many segments should be added to the \ac{mg}'s buffer in each scheduling slot}, which needs to satisfy two requirements, i.e., 1) buffer requirement: buffer minimum segments to avoid buffer length empty; 2) resource requirement: assign all bandwidth and computing resources to each \ac{smg} to observe how many segments can be buffered.

For the buffer requirement, we need to guarantee the buffer length is larger than the scheduling slot length for each \ac{smg}. Therefore, we can have
\begin{equation}
    {{n}_{\text{buffer}}}=\sum\limits_{g\in \mathcal{G}}{{{\left[ \frac{{{T}_\text{s}}-{{q}_{g}}}{\tau } \right]}^{+}}},
\end{equation}
where ${{T}_\text{s}}$ and $\tau $ represent the scheduling slot length and the segment length, respectively. {Here, $q_g$ represents \ac{smg} $g$'s buffer length for current watching video sequence and function ${{\left[ x \right]}^{+}}=\max \left\{ x,0 \right\}$, respectively.}

\begin{figure}[!t]
    \centering
    \includegraphics[width=\mysinglefigwidth]{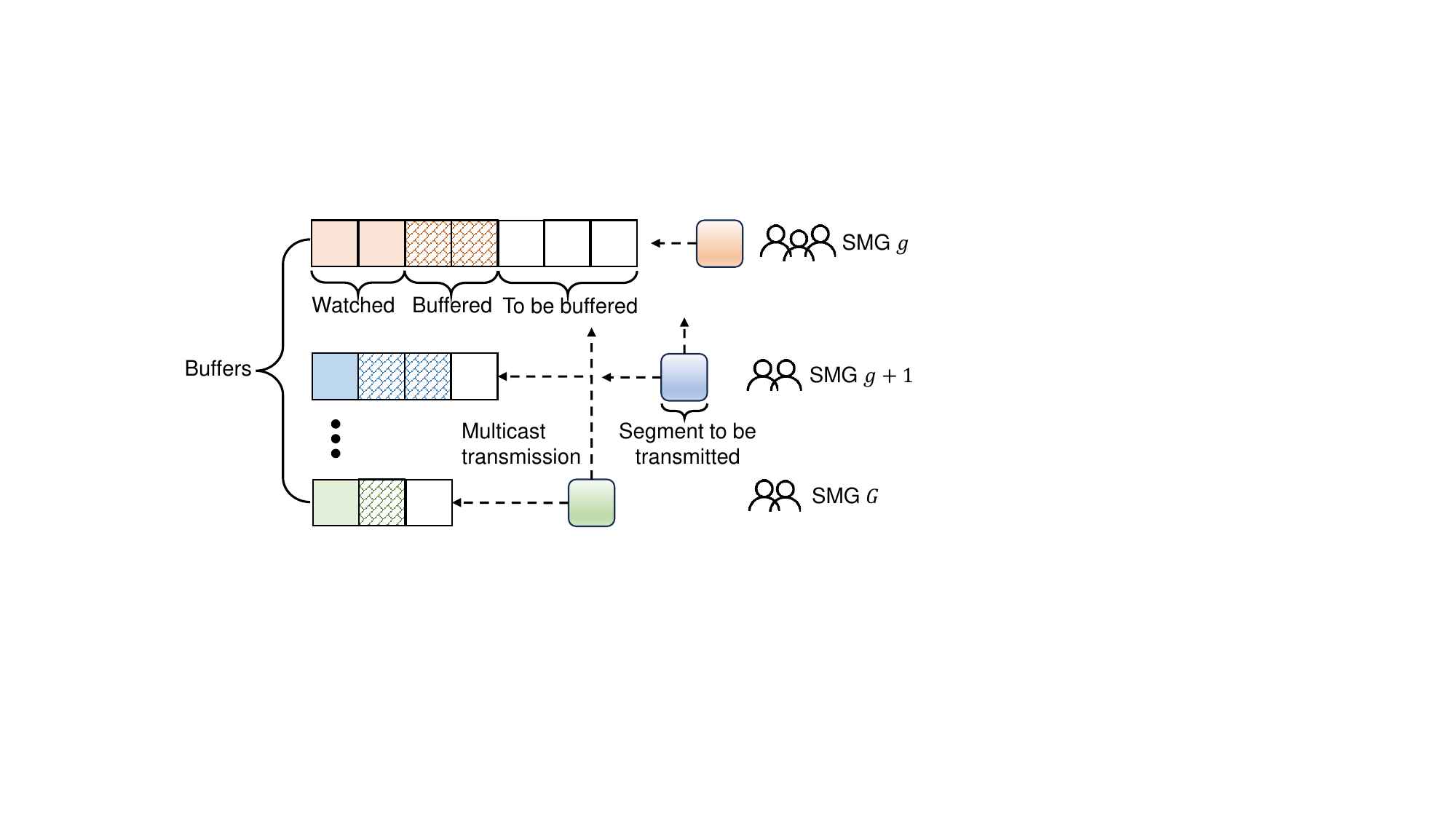}
    \caption{Multicast transmission mechanism among \ac{smg} groups.}
    \label{multi-trans}
\end{figure}

{For the resource requirement, we analyze the multicast transmission mechanism among \ac{smg} groups}, as shown in Fig.~\ref{multi-trans}. Each video sequence consists of multiple segments, represented by different colors. {The index orders of \acp{smg} are consistent with the video viewing positions, sorted from low to high.} Since video playback is continuous, the segment to be transmitted to \ac{smg} $g+1$'s buffer can also be transmitted to \ac{smg} $g$'s buffer to reduce repeated video transmission. Therefore, {the transmission capability of \ac{smg} $g+1$ needs to consider users' channel conditions both from itself and \ac{smg}~$g$.}
Based on the above analysis, the maximum buffered segments within all reserved bandwidth and computing resources need to satisfy the following requirements, i.e.,
\begin{align}
\label{const}
\begin{cases}
\sum\limits_{m=1}^{\tilde{n}_{g}^{\text{B}}}{\sum\limits_{l=1}^{{\bar{l}}}{z_{g,m}^{l}}}\le \underset{k\in \bigcup\nolimits_{d=1}^{g}{{{\mathcal{K}}_{d}}}}{\mathop{\min }}\,{{T}_\text{s}{r}_{g,k}}\le \sum\limits_{m=1}^{\tilde{n}_{g}^{\text{B}}+1}{\sum\limits_{l=1}^{{\bar{l}}}{z_{g,m}^{l}}},\forall g\in \mathcal{G}, \\ 
\mu \sum\limits_{m=1}^{\tilde{n}_{g}^{\text{C}}}{\sum\limits_{l=2}^{{\bar{l}}}{z_{g,m}^{l}}}\le {{T}_\text{s}}C\le \mu \sum\limits_{m=1}^{\tilde{n}_{g}^{\text{C}}+1}{\sum\limits_{l=2}^{{\bar{l}}}{z_{g,m}^{l}}},\forall g\in \mathcal{G}, \\ 
\end{cases}
\end{align}
where $z_{g,m}^{l}$ is the file size of segment $m$ of version $l$ for \ac{smg} $g$, which adopts the \ac{svc} principle~\cite{4317642}. Here, {$\tilde{n}_{g}^{\text{B}}$ and $\tilde{n}_{g}^{\text{C}}$ represent the segment buffering numbers by allocating reserved bandwidth resources and computing resources, respectively.} Here, $\bar{l}$ is the average segment version and parameter $\mu$ is the computing density for segment transcoding. {Here, $C$ is the computing capacity of \ac{ap}.} Furthermore, ${{r}_{g,k}}$ is the data rate for user $k$ in \ac{smg} $g$, which is given by
\begin{equation}
    r_{g,k}^{{}}=B{{\log }_{2}}\left( 1+\frac{{{\left| h_{g,k}^{{}} \right|}^{2}}{{P}_{\text{D}}}}{{{N}_{0}}} \right),
\end{equation}
where ${{h}_{g,k}}$ is the channel gain of user $k$ in \ac{smg} $g$ and ${{N}_{0}}$ is the noise power. Here, ${{P}_{\text{D}}}$ and $B$ represent the downlink transmission power and the reserved bandwidth resources for the \ac{mg}, respectively.

Since Eq.~\eqref{const} consists of monotonically increasing functions, {the maximum segment buffering numbers $\tilde{n}_{g}^{\text{B}}$ and $\tilde{n}_{g}^{\text{C}}$ can be uniquely determined by increasing their own values.} To satisfy both bandwidth and computing resource requirements, we select the smaller one between $\tilde{n}_{g}^{\text{B}}$ and $\tilde{n}_{g}^{\text{C}}$ as the segment buffering number, i.e., ${{\tilde{n}}_{g}}=\min \left\{ \tilde{n}_{g}^{\text{B}},\tilde{n}_{g}^{\text{C}} \right\}$. For one \ac{mg}, the segment buffering number in the resource requirement is the maximum value of all \acp{smg}' segment buffering numbers, i.e., ${{n}_{\text{resource}}}=\underset{g\in \mathcal{G}}{\mathop{\max }}\,{{\tilde{n}}_{g}}$. 

According to the buffer and resource requirements, we can determine the segment buffering number, i.e., ${{n}}=\lfloor\max \left\{ {{n}_{\text{buffer}}},{{n}_{\text{resource}}} \right\}\rfloor$, where $\lfloor \cdot \rfloor$ is the floor operation. Then, the buffered segment sequence of \ac{smg} $g$ can be obtained by integrating segment buffering number and order, denoted by ${{\Omega }_{g}}$.

\subsection{DT-Assisted Buffer Update Scheme}
\begin{figure}[!t]
    \centering
    \includegraphics[width=0.86\mysinglefigwidth]{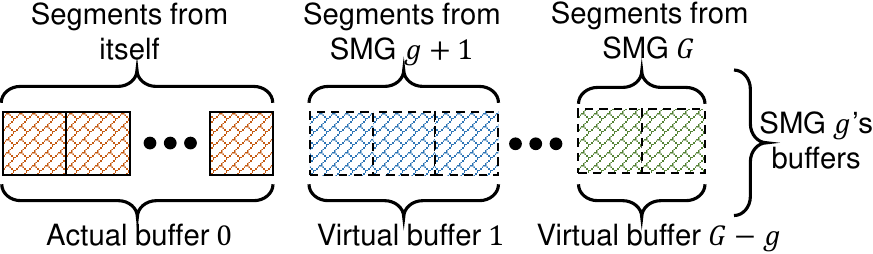}
    \caption{DT-assisted \ac{smg}'s buffer management.}
    \label{buffer}
\end{figure}
As shown in Fig.~\ref{buffer}, {we utilize the virtualization technology of DT to construct multiple virtual buffers for \ac{smg} $g$}, {where each virtual buffer corresponds to the divided and subsequent \ac{smg}'s buffer.} The buffer index of \ac{smg} $g$ is denoted by $f$, ranging from $0$ to $G-g$. In each time slot $t$, \ac{smg} $g$'s buffers are updated as 
\begin{align}
q_{g,t+1}^{f}=
\begin{cases}
{{\left[ q_{g,t}^{f}-{{T}_{s}}+\tau \left| {{\Omega }_{g}} \right| \right]}^{+}},&f=0, \\ 
q_{g,t}^{f}+\tau \left| {{\Omega }_{g+f}} \right|,&\forall f\in \left[ 1,G-g \right],
\end{cases}
\end{align}
where $\left| {{\Omega }_{g}} \right|$ is the segment buffering number of \ac{smg} $g$. 

When \ac{smg} $g$ starts to watch the next video, the current buffers $0$ to $G-1$ will be replaced by new buffers $1$ to $G$, and a new empty virtual buffer will be added as buffer $G$. Assuming that within each scheduling slot, the number of swipe behaviors for each \ac{smg} is at most once. An indicator function ${{\mathbb{S}}_{g,t}}$ is introduced, where ${{\mathbb{S}}_{g,t}}=1$ denotes a swipe behavior and ${{\mathbb{S}}_{g,t}}=0$ denotes no swipe behavior. By integrating the impact of swipe behaviors, \ac{smg} $g$'s buffer update can be further modified as
\begin{align}
\widetilde{q}_{g,t+1}^{f}=
\begin{cases}
q_{g,t+1}^{f},&{{\mathbb{S}}_{g,t}}=0, \\ 
q_{g,t+1}^{f+1},&{{\mathbb{S}}_{g,t}}=1 \& f\in \left[ 0,G-g-1 \right], \\ 
0,&{{\mathbb{S}}_{g,t}}=1 \& f=G-g. \\ 
\end{cases}
\end{align}

\section{Multicast QoE Model Establishment}\label{mqoe}
To evaluate the system performance, {we construct the multicast QoE model, consisting of rebuffering time, video quality, and quality variation, which considers the mutual influence of multicast segment buffering.} The specific analysis is presented as follows.

In each time slot, {since each \ac{smg} can occupy the total bandwidth and computing resources in its scheduling slot, we further divide a scheduling slot into multiple mini-slots}, where each \ac{smg} can occupy the total bandwidth and computing resources in its mini-slot. We define the variable $\beta_{g,t}$ as the division ratio for \ac{smg} $g$ in scheduling slot $t$, which need to satisfy the following constraint, i.e.,
\begin{equation}\label{fir}
    \sum\limits_{g=1}^{G}{{{\beta }_{g,t}}}\le 1,\forall t\in \mathcal{T}.
\end{equation}
where $\mathcal{T}$ is the scheduling slot set. For the simplification of expression, we omit $t$ in the following section. Since the lagging \ac{smg} can receive segments from the leading \ac{smg} in its mini-slot, the transmission delay of leading \ac{smg} needs to consider users' channel conditions of itself and its previous \acp{smg}. Correspondingly, the multicast transmission delay, $D_g$, is derived by
\begin{equation}
{{D}_{g}}=\frac{\sum\nolimits_{l=1}^{L}{\sum\nolimits_{m\in {{\Omega }_{g}}}{a_{g,m}^{l}\sum\nolimits_{j=1}^{l}{z_{g,m}^{j}}}}}{\underset{k\in \bigcup\nolimits_{d=1}^{g}{{{\mathcal{K}}_{d}}}}{\mathop{\min }}\,{{\beta }_{g}}r_{g,k}^{{}}}.
\end{equation}

Since each segment can have multiple layers that correspond to different versions, we define a binary version selection variable $a_{g,m}^{l}$, {where $a_{g,m}^{l}=1$ indicates segment layer $l$ of video $m$ is selected for buffering in \ac{smg} $g$}, otherwise, $a_{g,m}^{l}=0$. 

To avoid repeated video transmission, we assume that only one segment version can be selected for buffering in \ac{smg} $g$ in each time slot, which can be expressed as
\begin{equation}
    \sum\nolimits_{l=1}^{L}{a_{g,m}^{l}}=1, \forall m\in {{\Omega }_{g}}.
\end{equation}

Since the transmission process and transcoding process can be conducted in parallel, the service delay of \ac{smg} $g$ can be expressed as
\begin{equation}
    {{S}_{g}}=\max \left\{ {{D}_{g}},\frac{\mu \sum\nolimits_{l=1}^{L}{\sum\nolimits_{m\in {{\Omega }_{g}}}{a_{g,m}^{l}\sum\nolimits_{j=1}^{l}{z_{g,m}^{j}}}}}{{{\beta }_{g}}C} \right\}.
\end{equation}

Based on \ac{smg} $g$'s current actual buffer $0$ and service delay, we refer to \cite{8260530} to derive the rebuffering time, i.e.,
\begin{equation}
    {{R}_{g}}={{\left[ {{S}_{g}}-\widetilde{q}_{g}^{0} \right]}^{+}}.
\end{equation}

\ac{ssim} is a common metric used in video quality assessment. The relationship between video bitrate, $b$, and \ac{ssim}, $Q$, can be depicted as $Q=1-\frac{1}{2b+1}$~\cite{huang2020online}, where $Q$ ranges from 0 to 1 and a higher $Q$ means a higher video quality. Based on this mathematical relationship, the video quality of new buffered segments in \ac{smg} $g$ can be expressed as
\begin{equation}
{{Q}_{g}}=\sum\limits_{m\in {{\Omega }_{g}}}{1-\frac{1}{2\sum\nolimits_{l=1}^{L}a_{g,m}^{l}\sum\nolimits_{j=1}^{l}{z_{g,m}^{j}}/\tau+1}}.
\end{equation}

Based on video quality, we can analyze the video quality variation, $V_g$,
between adjacent segments in \ac{smg} $g$, depicted by
\begin{equation}
    {{V}_{g}}=\frac{1}{\left| {{\Omega }_{g}} \right|}\sum\limits_{m=1}^{\left| {{\Omega }_{g}} \right|}{\left| {{Q}_{g,m}}-{{Q}_{g,m-1}} \right|}.
\end{equation}
{where $Q_{g,0}$ is the video quality of the last segment in \ac{smg} $g$'s buffer before new segment buffering.} 

{To reflect the user's satisfaction with network management}, {three factors including rebuffering time, video quality, and quality variation}, can be integrated into the multicast QoE model, as referred in~\cite{zhou2022pdas}, which can be expressed as
\begin{equation}
    {{\Upsilon }_{g}}\left( {{ \bm{a}_{g}},{\beta }_{g}} \right)={{Q}_{g}}-\lambda_{{g,1}}{{R}_{g}}-\lambda _{g,2}{{V}_{g}},
\end{equation}
where $\bm{a}_{g}$ is a vector whose unit element is $a_{g,m}^{l}$. Here, $\lambda _{g,1}$ and $\lambda _{g,2}$ represent users' sensitivity degrees of rebuffering time and quality variation in \ac{smg} $g$, respectively. 

Considering each \ac{smg} has multiple segments to be buffered with different buffering orders, we need to incorporate the impact of buffering order into resource allocation. Therefore, we transform the buffering order into weighting factors, which can be expressed as
\begin{equation}
    {{\omega }_{g}}=\frac{\sum\nolimits_{m=1}^{{{\Omega }_{g}}}{{{\phi }_{g,m}}}}{\sum\nolimits_{g=1}^{G}{\sum\nolimits_{m=1}^{{{\Omega }_{g}}}{{{\phi }_{g,m}}}}},
\end{equation}
where $\phi_{g,m}$ is the buffering order of segment $m$ for \ac{smg} $g$, and a larger value corresponds to a higher buffering priority. Based on weighting factors, we can refine the established QoE model, i.e.,
\begin{equation}
    {{\widetilde{\Upsilon }}_{g}}=\omega_{g}\Upsilon_{g}.
\end{equation}

\section{Problem Formulation and Solution}\label{problem}

\subsection{Problem Formulation}
{Since each \ac{smg}'s bandwidth and computing resource demands are dynamic due to users' dynamic swipe behaviors and video requests, the reserved bandwidth and computing resources need to be flexibly and accurately allocated to each \ac{smg} at each scheduling slot to reduce playback lags.} Furthermore, the versions of segments to be buffered need to be as high and stable as possible to ensure high video playback quality. By achieving low playback lags and high video playback quality, users can obtain better \ac{QoE}. Based on the above analysis, our objective is to maximize users' long-term QoE by optimizing segment version selection and slot division at each scheduling slot. The formulated problem \textbf{P1} is given by

    \begin{align}
   \textbf{P1}:&\underset{{{\left\{ {\bm{a}_{g,t}},{{\beta }_{g,t}} \right\}}_{t\in \mathcal{T}}}}{\mathop{\max }}\,\frac{1}{T}\sum\nolimits_{t=1}^{T}{\sum\nolimits_{g\in \mathcal{G}}{\widetilde{\Upsilon} }_{g}}\left( {{ \bm{a}_{g,t}},{\beta }_{g,t}} \right) \\ 
  \text{s.t.} \; &(9) \;\text{and}\; (11),\notag \\ 
 & a_{g,m,t}^{l}\in \left\{ 0,1 \right\},\forall g\in \mathcal{G},m\in {{\Omega }_{g}},l\in \mathcal{L},t\in \mathcal{T},\tag{19a} \\ 
 & {{\beta }_{g,t}}\in [0,1],\forall g\in \mathcal{G},t\in \mathcal{T}.\tag{19b} 
\end{align}
Constraint (9) is the resource capacity constraint, which guarantees that the total scheduled bandwidths and computing resources cannot exceed the system capacity. Constraint (11) is the video transmission constraint, which avoids the repeated video transmission for each \ac{smg}.

\subsection{Solution}
\begin{figure}[!t]
    \centering
    \includegraphics[width=\mysinglefigwidth]{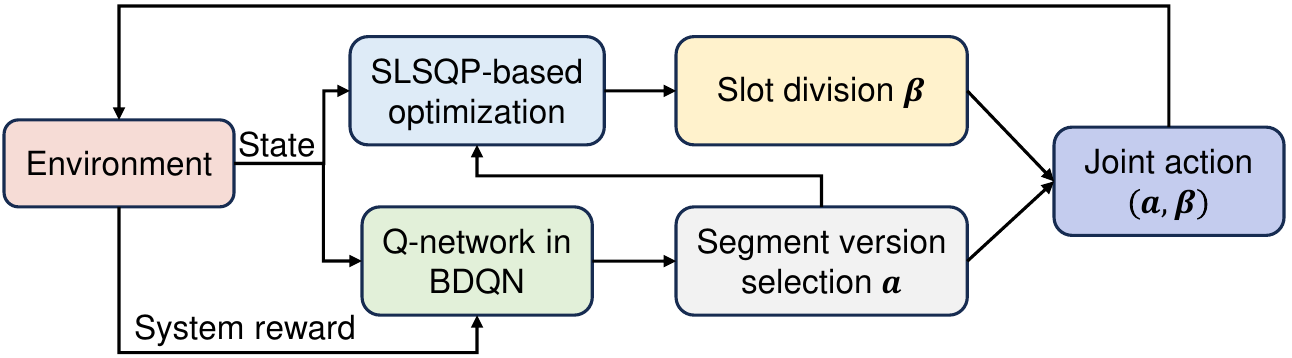}
    \caption{The proposed data-model-driven algorithm structure.}
    \label{joint}
\end{figure}

{We omit $t$ in this subsection for the simplification of expression.} The formulated problem is a mixed-integer nonlinear programming problem with the objective of maximizing users' long-term QoE. {The dimension of $\bm{a}$ is ${{2}^{L\sum\nolimits_{g\in \mathcal{G}}{g\left| {{\Omega }_{g}} \right|}}}$, which is very huge when the numbers of \acp{smg}, recommended segments, and segment versions are high.} {The variable vector $\bm{\beta}$ are continuous values ranging from 0 to 1.} Since these variables are coupled with each other, it is hard to directly use a data-driven or model-based method to solve it~\cite{10373769,10092806}. Therefore, we consider a data-model-driven algorithm to solve it, as shown in Fig.~\ref{joint}. The state from the environment is first input into Q-network in the \ac{bdq}~\cite{tavakoli2018action} to solve the high dimension of variable $\bm{a}$. Based on the segment version selection, we then utilize the \ac{slsqp} algorithm to obtain the optimal $\bm{\beta}$. Finally, the joint action $\{\bm{a},\bm{\beta}\}$ is fed back to the environment to obtain the system reward, which is utilized for Q-network update. The detailed procedure is presented as follows. 

\subsubsection{DRL-Based Segment Version Selection}

\begin{figure}[!t]
    \centering
\includegraphics[width=\mysinglefigwidth]{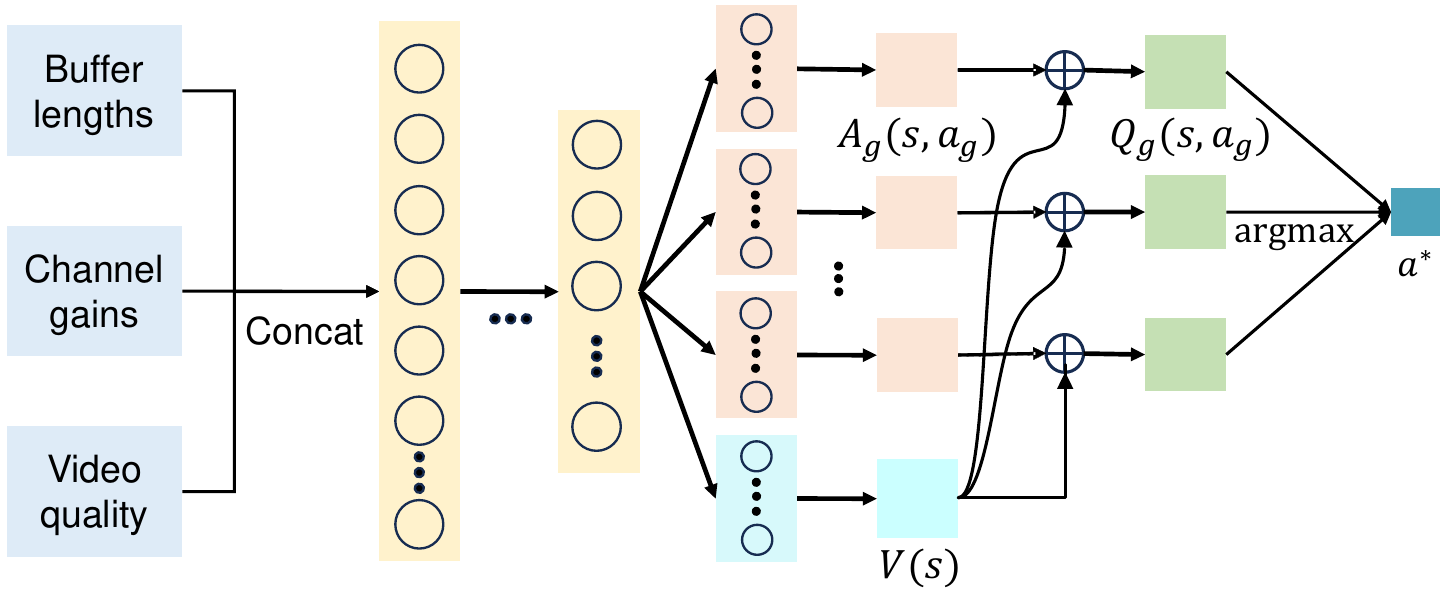}
    \caption{The constructed BDQN architecture.}
    \label{BDQ-A}
\end{figure}

To solve the segment version selection optimization subproblem, we employ the \ac{bdq} algorithm, whose architecture is shown in Fig.~\ref{BDQ-A}.

As shown in Fig.~\ref{BDQ-A}, \acp{smg}' state consisting of buffer lengths, channel gains, and video quality, i.e., $s=\left\{ {{\left\{ \tilde{q}_{g}^{0} \right\}}_{g\in \mathcal{G}}}, {{\left\{ {{h}_{g,k}} \right\}}_{k\in {{\mathcal{K}}_{g}},g\in \mathcal{G}}},{{\left\{ {{Q}_{g,0}} \right\}}_{g\in \mathcal{G}}}\right\}$, are concatenated and input into the fully connected neural networks. The BDQN architecture is an amalgamation of Dueling Q-Networks~\cite{wang2016dueling} and a branching action mechanism, which comprises two key components, i.e., value function and advantage function. The value function $V(s)$ estimates the value of being in a particular state, independent of any specific action. {Advantage function $A_g$ exists in each action branch $g$, which estimates the additional value of taking a particular action in a given state, relative to other actions.} The segment version selection decision is split into $G$ sub-actions, where the sub-action for \ac{smg} $g$ is denoted by ${\bm{a}_{g}}={{\left\{ a_{g,m}^{l} \right\}}_{m\in {{\Omega }_{g}},l\in \mathcal{L}}}$. Each action branch corresponds to a \ac{smg}. The reward function is \acp{smg}' QoE, i.e., $\sum\nolimits_{g\in \mathcal{G}}{{{\widetilde{\Upsilon }}_{g}}\left( {\bm{a}_{g}},{{\beta }_{g}} \right)}$. The advantage function of each sub-action, i.e., ${{A}_{g}}(s,{\bm{a}_{g}})$ is trained with the common state value $V(s)$ by experience replay. The Q-value of each sub-action is updated based on the average advantage functions, which can be expressed as
\begin{equation}
{{\mathbb{Q}}_{g}}\left( s,{\bm{a}_{g}} \right)=V\left( s \right)+\left( {{A}_{g}}\left( s,{{a}_{g}} \right)-\frac{1}{\rho_g}\sum\limits_{\bm{a}_{g}^{'}\in {{A}_{g}}}{{{A}_{g}}\left( s,\bm{a}_{g}^{'} \right)} \right),    
\end{equation}
where $\bm{a}_g^{'}$ is the next-step sub-action for \ac{smg} $g$ and $\rho_g$ is the sub-action dimension, i.e., $2\times\|\Omega_g\|\times L$. {In each step, BDQN has a probability, i.e., $\left( 1-\varsigma  \right)$, to select the action that can obtain the highest Q-value}, which can be expressed as
\begin{equation}\label{act}
    \bm{a}=\left\{ \underset{\bm{a}_{1}^{'}}{\mathop{\arg \max }}\,{{\mathbb{Q}}_{1}}\left( s,\bm{a}_{1}^{'};{\bm{\theta }_{1}} \right),\cdots ,\underset{\bm{a}_{G}^{'}}{\mathop{\arg \max }}\,{{\mathbb{Q}}_{G}}\left( s,\bm{a}_{G}^{'};{\bm{\theta }_{G}} \right) \right\},
\end{equation}
where $\bm{\theta}_g$ is the network weights of Q-network $g$. 

To enable the agent to efficiently train, {we employ the temporal-difference (TD) target after every step}, which is an estimate of the expected return (future cumulative reward) for a given state-action pair, expressed by 
\begin{equation}\label{td-error}
    y=r+\vartheta \frac{1}{G}\sum\limits_{g}{{{\widehat{\mathbb{Q}}}_{g}}\left( s',\underset{\bm{a}_{g}^{'}}{\arg \max} {\mathbb{Q}_{g}}\left( s', \bm{a}_{g}^{'} \right) \right)}.
\end{equation}

To quantify the difference between the predicted and target Q-values, {a loss function is essential, which can guide the optimization of neural network parameters to improve learning accuracy.} The loss is the expected value of mean square error across the branches, i.e.,
\begin{equation}\label{loss}
    L\left( \bm{\theta}  \right)={{\text{E}}_{\left( s,\bm{a},r,s' \right)\sim \mathcal{D}}}\left[ \frac{1}{G}\sum\limits_{g}{{{\left( {{y}}-{\mathbb{Q}_{g}}\left( s,{\bm{a}_{g}} \right) \right)}^{2}}} \right].
\end{equation}

Furthermore, the prioritization error is crucial for efficiently focusing the learning process on the most informative experiences, by prioritizing those with higher TD errors in the training process. Here, the prioritization error is defined by summing across a transition's absolute, i.e., 
\begin{equation}\label{pri}
    {{\delta }_{\mathcal{D}}}\left( s,\bm{a},r,s' \right)=\sum\limits_{g}{\left| {{y}_{g}}-{\mathbb{Q}_{g}}\left( s,{\bm{a}_{g}} \right) \right|}.
\end{equation}

\begin{algorithm}[t]\label{alg1}
\caption{BDQN-based segment version selection}
\textbf{Input:} \acp{smg}' buffer lengths $\widetilde{\bm{q}}$, channel gains $\bm{h}$, video quality $\bm{Q}$, and network update threshold $\Gamma$;

\textbf{Output:} Segment version selection $\bm{a}$;

\textbf{Initialize:} Replay memory $\mathcal{D}$, action-value function $\mathbb{Q}$ with random weights $\bm{\theta}$, target action-value function $\widehat{\mathbb{Q}}$ with weights $\bm{\theta}^{'}$;

\For{each $\text{episode}$}
    {
    Reset initial state $s_1$;
    
        \For{\text{each step} $t \in \{1,\cdots,T\}$}
            {
            With probability $\varsigma$ select a random action $\bm{a}_t$,
            otherwise, the action is selected based on Eq.~\eqref{act};

            Implement the Algorithm~\ref{alg2} to obtain slot division $\bm{\beta}_t$;
            
            Execute joint action $(\bm{a}_t,\bm{\beta}_t)$ in the environment;
            
            Observe reward $r_t$ and new state $s_{t+1}$;
            
            Store transition $(s_t, \bm{a}_t, r_t, s_{t+1})$ in $\mathcal{D}$;
            
            Prioritize replay based on Eq.~\eqref{pri} to obtain a transition $(s_j, a_j, r_j, s_{j+1})$ from $\mathcal{D}$;
            
            Calculate the TD target based on Eq.~\eqref{td-error};
            
            Perform a gradient descent step based on the loss in Eq.~\eqref{loss};
            
            Every $\Gamma$ steps, reset $\widehat{\mathbb{Q}} = \mathbb{Q}$;
            }
    }
\end{algorithm}

Based on the above analysis, the detailed algorithm procedure to determine segment version selection variable $\bm{a}$ is shown in Algorithm~\ref{alg1}.

\subsubsection{Optimization-Based Slot Division}
We first transform the maximization problem into the minimization problem. When $\bm{a}$ is determined, denoted by $\bm{a}^{*}$, the original objective function in the maximization problem can be transformed to the opposite objective function in the minimization problem, i.e., $\sum\nolimits_{g\in \mathcal{G}}\widehat{\Upsilon}_g\left( {\bm{a}_{g}}^{*},{{\beta }_{g}} \right) = -\sum\nolimits_{g\in \mathcal{G}}\widetilde{\Upsilon }_{g}^{{}}\left( {\bm{a}_{g}}^{*},{{\beta }_{g}} \right)$.

\begin{theorem}
\label{theo1}
The transformed objective function $\sum\nolimits_{g\in \mathcal{G}}\widehat{\Upsilon }_{g}^{{}}\left( {\bm{a}_{g}}^*,{{\beta }_{g}} \right)$ is convex about $\bm{\beta}$.
\end{theorem}
\begin{IEEEproof}
See Appendix~\ref{app:theo1}.
\end{IEEEproof}

For the transformed convex problem, we need to find an effective algorithm to solve it. \ac{slsqp} is adept at solving the formulated linear constrained convex optimization problem by iteratively approximating them into quadratic subproblems~\cite{kraft1988software}. The algorithm can converge to a global minimum for a convex problem due to the following reasons. First, {by leveraging the gradient information and incorporating the constraints into Lagrange multipliers}, \ac{slsqp} ensures that the solution not only optimizes the objective function but also strictly adheres to the problem's constraints. Second, given that our decision variables are continuous and bounded, \ac{slsqp}'s design inherently aligns with the problem's structure, making it reliable to find the optimal solution. \ac{slsqp} needs the gradient information, but the optimization problem is not derivative at every point. Therefore, we employ the sub-gradient method \cite{boyd2003subgradient} to analyze its sub-gradient.

In iteration $i$, the sub-gradient of $\widehat{\Upsilon} _{g}\left( {\bm{a}_{g}}^{*},{{\beta }_{g}^{(i)}} \right)$ can be determined based on the value of ${{(x)}^{+}}$ function. Let denote $\widehat{\Upsilon} _{g}\left( {\bm{a}_{g}}^{*},{{\beta }_{g}^{(i)}} \right)=\frac{1}{\beta _{g}^{(i)}}\varphi _{g,1}-\varphi _{g,2}$, where functions $\varphi_{g,1}$ and $\varphi_{g,2}$ can be expressed as
$\varphi_{g,1}={{\omega }_{g}}\lambda _{g,1}\max \left\{\Xi_1(\beta_g),\Xi_2(\beta_g) \right\}$ and $\varphi_{g,2}=\widetilde{q}_{g}^{0}(\bm{a}_g^*)$. Functions $\varphi _{g,1}$ and $\varphi_{g,2}$ are constants related with ${\bm{a}_{g}}^{*}$.
In addition, functions $\Xi_1(\beta_g)$ and $\Xi_2(\beta_g)$ are defined in Appendix A.

Then, we can have the sub-gradient of variable for slot division variable 
$\beta_g^{(i)}$:
\begin{align}
\nabla \widehat{\Upsilon }(\beta _{g}^{(i)})=
\begin{cases}
-\frac{\varphi _{g,1}}{{{\left( \beta _{g}^{(i)} \right)}^{2}}},&\beta _{g}^{(i)}<\frac{\varphi _{g,1}}{\varphi _{g,2}}, \\  
-\sigma \frac{\varphi _{g,1}}{{{\left( \beta _{g}^{(i)} \right)}^{2}}},&\beta _{g}^{(i)}=\frac{\varphi _{g,1}}{\varphi _{g,2}}, \\ 
0,&\text{otherwise},
\end{cases}
\end{align}
where $\sigma$ is a positive value with the range of $(0, 1]$.

Based on the sub-gradient information, the Hessian matrix, $\myvec{H}$, is expressed as $\mathtt{diag}\left( \nabla^{2} \widehat{\Upsilon}\left( \beta _{1}^{(i)} \right),\ldots ,\nabla^{2} \widehat{\Upsilon} \left( \beta _{G}^{(i)} \right) \right)$. Since the quadratic subproblem is a fundamental step in the \ac{slsqp} algorithm, we formulate the quadratic subproblem as
\begin{align}
   \textbf{P2:} &\underset{\bm{\beta}^{(i)}}\min \widehat{\Upsilon }\left( {\bm{a}^{*}},{\bm{\beta }^{(i)}} \right)+\nabla \widehat{\Upsilon }{{\left( {\bm{a}^{*}},{\bm{\beta }^{(i)}} \right)}^{T}}\myvec{d}+\frac{1}{2}{{\myvec{d}}^{T}}\myvec{H}\myvec{d} \\ 
 \text{s.t.}\;  & c\left( {\bm{\beta }^{(i)}} \right)+\nabla c{{\left( {\bm{\beta }^{(i)}} \right)}^{T}}\myvec{d}\le 0, \tag{26a} \\ 
 & {\bm{\beta }^{(i)}}+\myvec{d}\in \left[ 0,1 \right], \tag{26b}
\end{align}
where the objective function is the quadratic approximation around $(\bm{a}^*,\bm{\beta}^{(i)})$. Here, $c$ and $\myvec{d}$ represent the function of Eq.~\eqref{fir} and the direction of change, respectively.

\begin{algorithm}[t]
	\caption{SLSQP-based slot division}
	\label{alg2}
	
	\textbf{Input:} Segment version selection $\bm{a}$, video bitrate sequence $\mathbf{z}$, bandwidth $B$, computing capacity $C$, downlink transmission power $P^{\text{DL}}$, noise power $N_0$, and buffer length $\mathbf{\widetilde{q}}$;
	
	\textbf{Output:} Optimal slot division $\bm{\beta}^*$;

 	\textbf{Initialize:} $i$ = 0, ${\bm{\beta }^{(i)}}={\bm{\beta }^{(0)}}$, converge = False;

    \While{converge == False}
    {
        Calculate the gradient $\nabla \widehat{\Upsilon }(\bm{\beta} _{{}}^{(i)})$;

        Approximate the objective function and constraints at $\left( {\bm{a}^{*}},{\bm{\beta }^{(i)}} \right)$ by quadratic functions;

        Formulate the quadratic subproblem \textbf{P2};

        Solve the quadratic subproblem to get direction $\myvec{d}^{(i)}$;

        Update the solution using a line search: ${\bm{\beta}^{(i+1)}}={\bm{\beta }^{(i)}}+\alpha \cdot {\myvec{d}^{(i)}}$;

    \If{$\left\| {\bm{\beta }^{(i+1)}}-{\bm{\beta }^{(i)}} \right\|<\varepsilon$}
    {
        converge = True;
    }

    Increase the iteration number, $i \leftarrow i+1$;
    }
    Return ${\bm{\beta }^{*}}={\bm{\beta }^{(i)}}$;
\end{algorithm}

To solve the formulated quadratic subproblem, we can utilize the QP solver in the CVXOPT\footnote{CVXOPT: https://cvxopt.org/examples/tutorial/qp.html}. The specific algorithm is shown in Algorithm~\ref{alg2}.

\section{Simulation Results}\label{results}

We conduct extensive simulations on the real-world dataset to evaluate the performance of the proposed DT-based network management scheme.

\subsection{Simulation Setup}


\begin{table}[!t]
\centering
\caption{Simulation Parameters}
\label{sim}
\begin{tblr}{
    width = 0.98\linewidth,
    colspec = {X[1.3,c,m]X[2,c,m]X[1.3,c,m]X[2,c,m]},
    columns = {mode=dmath},
    row{1} = {mode=text},
    row{1} = {font=\bfseries},
    hlines,
    hline{2} = {1}{-}{},
    hline{2} = {2}{-}{},
    vline{2,4},
    vline{3} = {1}{-}{},
    vline{3} = {2}{-}{},
}
    Parameter & Value & Parameter & Value \\
    B & [6, 14]~\myunit{MHz} & K & [10, 26] \\
    C & [8, 12]~\myunit{G cycles/s} & \mu & 4~\myunit{G cycles/Mb} \\
    T_\text{s} & 5~\myunit{sec}	 & \lambda_{1} & [0.2, 0.4] \\
    \tau	& 2~\myunit{sec}	& \lambda_{2}	& [0.5, 0.7] \\
    P_\text{D}	& 27~\myunit{dBm}	& N_0	& -174~\myunit{dBm}
\end{tblr}
\end{table}

\begin{figure}[!t]
    \centering
\includegraphics[width=\mysinglefigwidth]{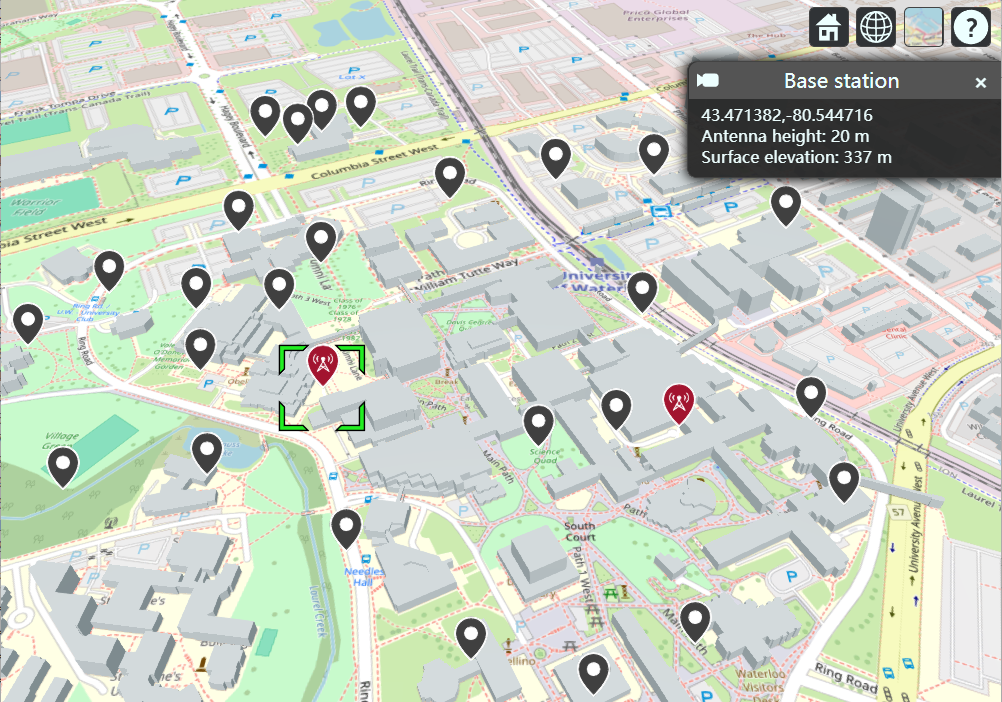}
    \caption{The simulation scene, where BSs and users are represented by red and gray icons, respectively.}
    \label{fig:BSs}
\end{figure}

We adopt the short video streaming dataset\footnote{ACM MM Grand Challenges: https://github.com/AItransCompetition/Short-Video-Streaming-Challenge/tree/main/data} to obtain users' swipe behaviors. We sample $1000$ short videos from the YouTube 8M dataset\footnote{YouTube 8M dataset: https://research.google.com/youtube8m/index.html}, which includes $8$ video types, i.e., Entertainment, Games, Food, Sports, Science, Dance, Travel, and News. Each video sequence is encoded into four versions, i.e., $L=4$. We consider the scenario where two BSs are deployed at the University of Waterloo (UW) campus and users' initial positions are randomly and uniformly generated around two BSs, as shown in Fig.~\ref{fig:BSs}. Each user moves along a prescribed path within the UW campus at a speed of $2\sim5~\myunit{km/h}$, and the corresponding channel path loss is obtained by the propagationModel at Matlab. The main simulation parameters are presented in Table \ref{sim}. 


Our BDQN architecture consists of four fully connected layers, transitioning from an initial state-size input to a layer with $128$ nodes. This architecture is split into two streams: a value stream with a single output and multiple advantage streams, which can produce a matrix of action advantage values that represents the combination of actions and their respective choices. To refine the learning process, we integrate a prioritized replay buffer, which emphasizes learning from experiences with higher predicted errors. The parameter setting and model structure are presented in Table~\ref{bdqn}.


\begin{table}[!t]
\centering
\caption{BDQN Parameters}
\label{bdqn}
\begin{tblr}{
    width = 0.98\linewidth,
    colspec = {X[1.5,c,m]X[1,c,m]X[1.5,c,m]X[1,c,m]},
    column{2,4} = {mode=dmath},
    row{1} = {mode=text},
    row{1} = {font=\bfseries},
    hlines,
    hline{2} = {1}{-}{},
    hline{2} = {2}{-}{},
    vline{2,4},
    vline{3} = {1}{-}{},
    vline{3} = {2}{-}{},
}
    Parameter & Value & Parameter & Value \\
    Memory size &	5000	  & Episode length	& 75 \\ 
    Initial epsilon	& 1	  & Discount factor 	& 0.9 \\ 
    Epsilon decay	& 0.99	  & Learning rate	& 0.001 \\
    Final epsilon	& 0.1	& Batch size	& 64 \\
    Number of episodes& 500	& NN layer connection	& \text{FC} \\
    Hidden layer structure & 512\times256\times256\times128 & Activation function & \text{ReLu} \\
    Advantage stream structure & 128\times36 & Value stream structure & 128\times1
\end{tblr}
\end{table}

\begin{figure}[!t]
    \centering
\includegraphics[width=0.8\mysinglefigwidth]{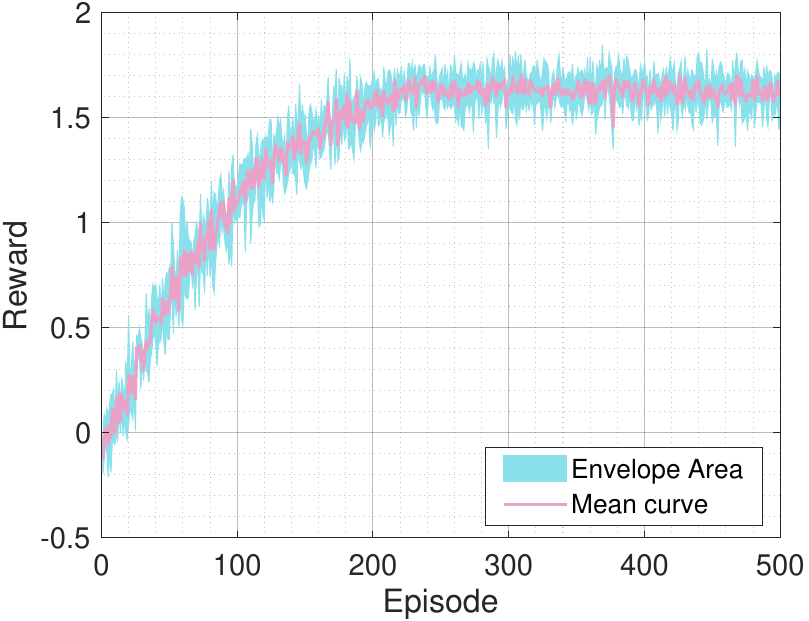}
    \caption{Convergence performance of proposed DT-based network management scheme.}
    \label{fig:converge}
\end{figure}

\begin{figure*}[t]
	\centering
	\subfloat[Buffer length comparison]{
		\includegraphics[width=0.305\textwidth]{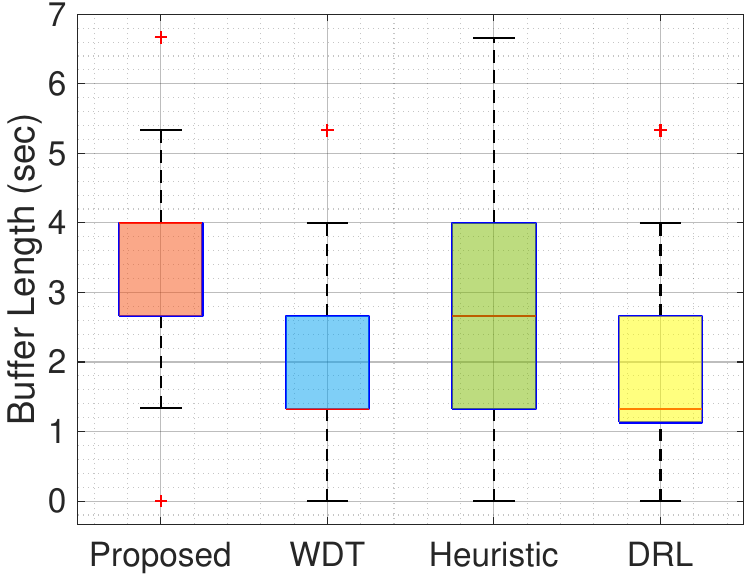}\label{10a}}
	\centering
	\subfloat[Video quality comparison]{
		\includegraphics[width=0.32\textwidth]{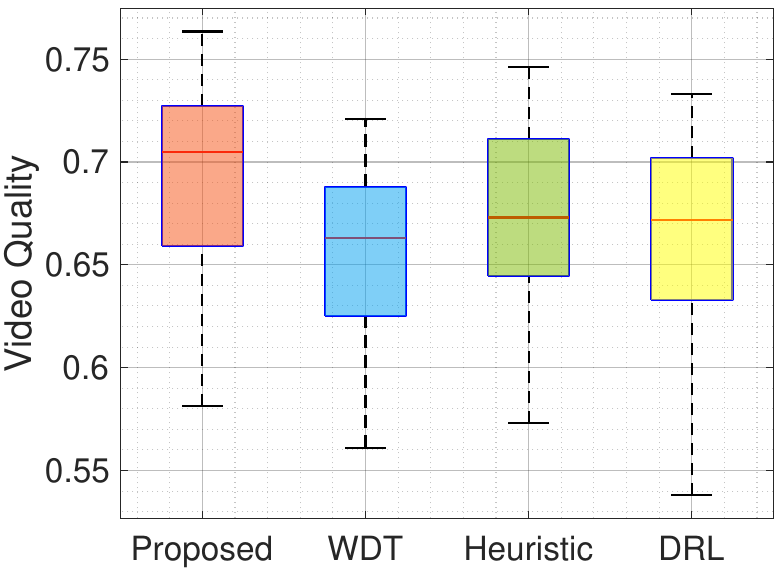}\label{10b}}
	\centering
	\subfloat[Video quality variation comparison]{
		\includegraphics[width=0.32\textwidth]{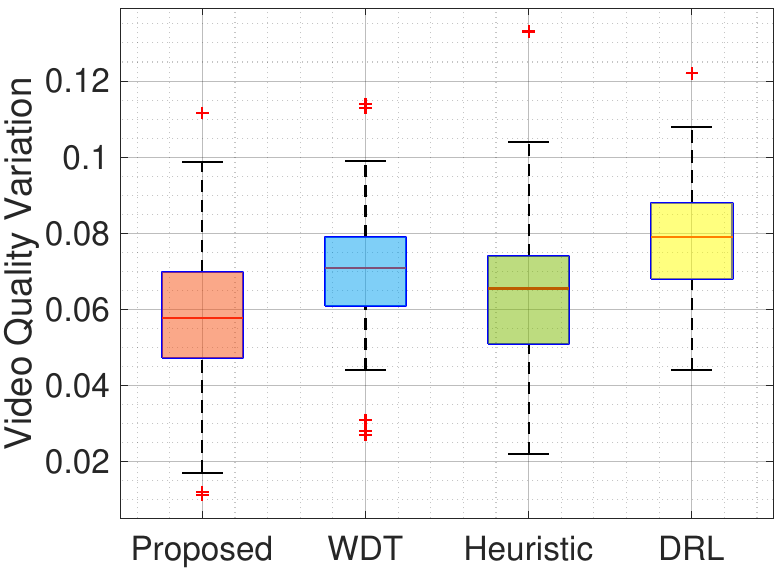}\label{10c}}
  \caption{The performance comparison of different QoE components.}
	\label{fig:components}
\end{figure*}

We compare the proposed DT-based network management scheme with the following benchmark schemes:
\begin{itemize}
    \item 
    \textbf{Without DT (WDT) scheme}: The segment buffering is based on the sequential principle. The rebuffering time estimation is based on the service delay and the \acp{smg}' currently total buffered segments. The determination of segment version selection and slot division employs the same data-model-driven method proposed in the DT-based network management scheme.
    \item 
    \textbf{Heuristic scheme}~\cite{9525340}: The segment buffering and buffer update employ the same principle proposed in the DT-based network management scheme. The scheduling slot is discretized into $10$ mini-slots. Each mini-slot is first provisionally allocated to each SMG, and then the corresponding segment version selection is determined by the branch and bound algorithm. Finally, the mini-slot is ultimately allocated to the SMG that can obtain the maximum QoE.
    \item
    \textbf{DRL-based scheme}: The segment buffering is based on the sequential principle. The rebuffering time estimation is based on the service delay and the \acp{smg}' currently total buffered segments. The joint optimization of segment version selection and slot division is determined by the modified DDPG algorithm~\cite{9764370}, where the range of segment version selection action is divided into four parts, i.e., $[0, 0.25)$, $[0.25, 0.5)$, $[0.5, 0.75)$, $[0.75, 1]$, corresponding to the segment version from low to high. Furthermore, a hierarchical reward is used to accelerate the algorithm's training speed. 
    
\end{itemize}

\subsection{Convergence Analysis}
In this subsection, we analyze the convergence performance of the proposed scheme within $500$ training episodes. As shown in Fig.~\ref{fig:converge}, we present the convergence curve of the DT-based network management scheme. We conducted four training trials to draw the corresponding envelope area and mean curve, where each training trail corresponds to a unique seed for action exploration and experience replay. Each episode consists of $75$ steps, and the corresponding reward is the average reward for all steps within an episode. It can be observed that as the number of episodes increases, the reward gradually grows larger. When the number of episodes approaches nearly $230$, the reward converges to a stable state, {indicating that the DT-based network management scheme can achieve a high and stable QoE for users.}

\subsection{Performance Evaluation of QoE Components}
In this subsection, we evaluate the performance of the \ac{mg}'s QoE components. The bandwidth, computing capacity, and user number are set to $10~\myunit{MHz}$, $10~\myunit{Gcycles/s}$, and $10$ respectively. {We present the box plot comparison of buffer length, video quality, and video quality variation across four different schemes in Fig.~\ref{fig:components}.} Each box plot delineates the interquartile range, median, and outliers. 

As shown in Fig.~\subref*{10a}, the proposed scheme demonstrates a higher median buffer length and a more compact interquartile range relative to the other schemes, which indicates users' equipment can buffer more segments to reduce the rebuffering probability. The compactness of the proposed scheme suggests lower variability in buffer length, which could bring a more stable user experience during video playback. The reason that the proposed scheme can achieve a better buffer length performance is attributed to the proposed DT-assisted segment buffer scheme, which can effectively abstract the watching probability distribution for priority-based buffering and maintain multiple virtual buffers for accurate buffer length updates. 

Fig.~\subref*{10b} presents a comparative analysis of video quality performance among different schemes with the range from $0.53$ to $0.77$. The proposed scheme reveals the highest median video quality, as well as a relatively narrower interquartile range compared to the other schemes. {This suggests that the proposed scheme not only delivers a relatively higher video quality but also ensures smoother playback in the quality of streaming content.} The reduced spread of data points and fewer outliers underscores its ability to provide a reliably high-quality watching experience, due to its advanced segment version selection algorithm that can achieve a good trade-off between buffering and streaming quality. 

As shown in Fig.~\subref*{10c}, we present the comparison of video quality variation. Based on the observation, the proposed scheme shows the lowest median value that suggests a central tendency towards lower variation in video quality. Despite the interquartile range of the proposed scheme being very close to the other schemes, the concentration of data around a lower median value and the reduced number of extreme outliers reflect its effectiveness in ensuring stable video quality. 

\subsection{Performance Evaluation of QoE Under Different Settings}
\begin{figure*}[!t]
	\centering
	\subfloat[QoE vs. different numbers of users]{
		\includegraphics[width=0.32\textwidth]{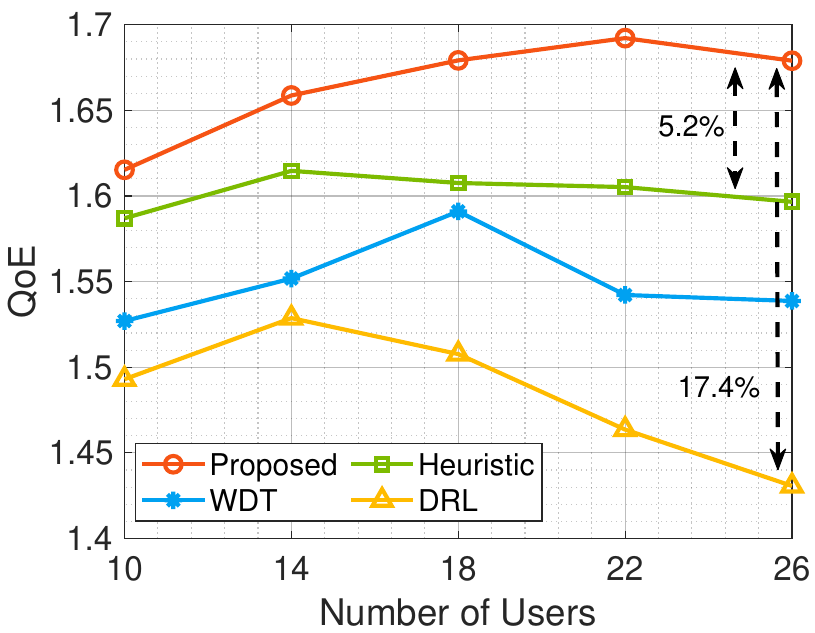}\label{11a}}
	\centering
	\subfloat[QoE vs. different bandwidths]{
		\includegraphics[width=0.315\textwidth]{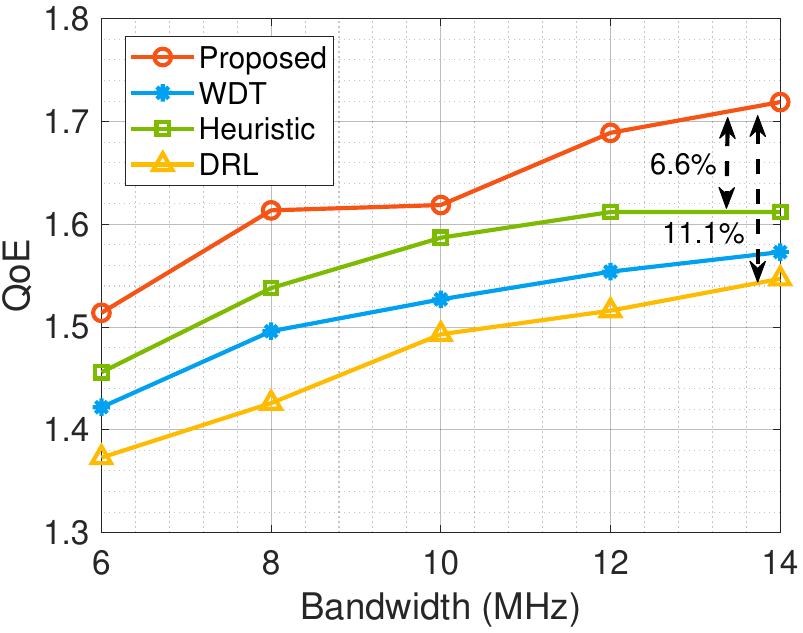}\label{11b}}
	\centering
	\subfloat[QoE vs. different computing capacities]{
		\includegraphics[width=0.315\textwidth]{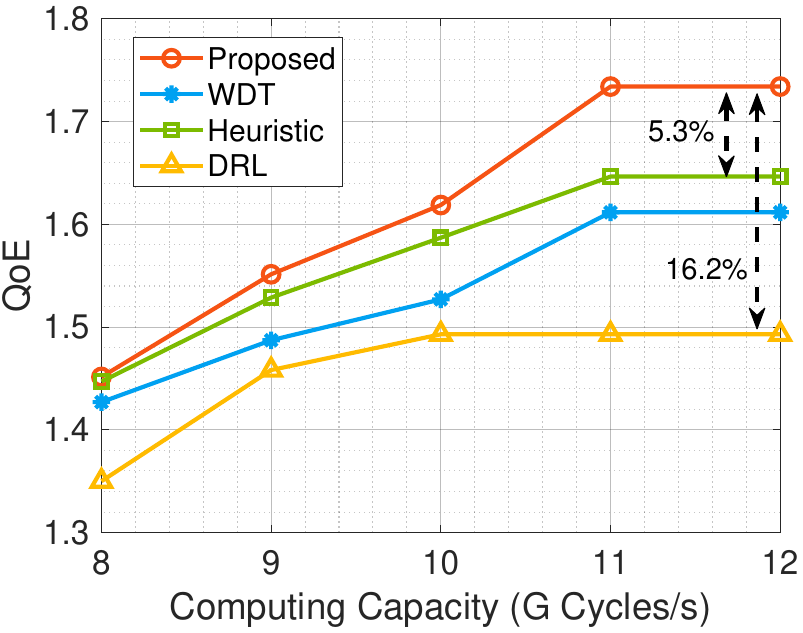}\label{11c}}
  \caption{QoE comparison under different users, bandwidths, and computing capacities.}
	\label{fig:metrics}
\end{figure*}

\begin{figure}[!t]
	\centering
	\subfloat[QoE vs. different sensitivity degrees of rebuffering time]{
		\includegraphics[width=0.38\textwidth]{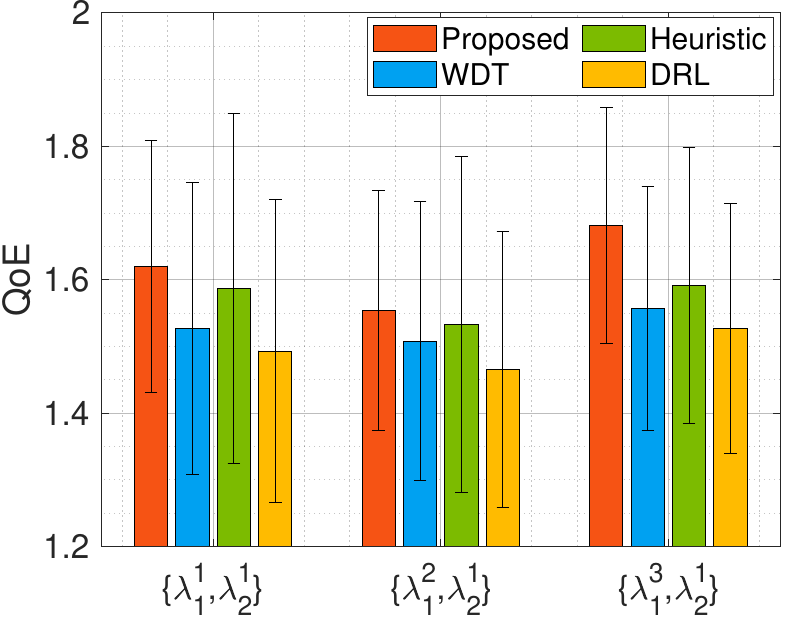}\label{12a}}
	\\
	\subfloat[QoE vs. different sensitivity degrees of video quality variation]{
		\includegraphics[width=0.38\textwidth]{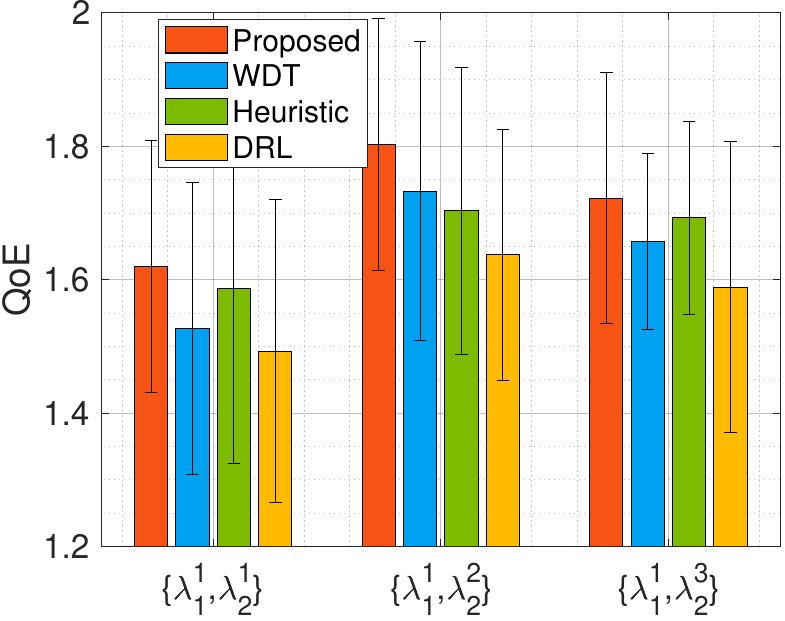}\label{12b}}
  \caption{QoE comparison under different sensitivity degrees.}
	\label{fig:lambda}
\end{figure}

We present QoE comparison under different users, bandwidths, and computing capacities among different schemes in Fig.~\ref{fig:metrics}. Overall, the proposed scheme can achieve superior performance under different simulation settings. 

Fig.~\subref*{11a} describes the correlation between the number of users and QoE. Based on the observation, {we find that QoE initially increases and then decreases as the number of users increases due to the mutual influence of multicast segment buffering.} Specifically, the increased users can lead to more users being clustered into one \ac{smg} and enrich users' diversity. Initially, the same multicast segments watched by the original users may gain a better QoE for newly clustered users who have low sensitivity degrees. However, with the increased number of users, the multicast segments cannot always satisfy users' differentiated watching requirements, which finally leads to low QoE. The proposed scheme exhibits a QoE of approximately $1.68$ when catering to $26$ users, a $5.2\%$ and $17.4\%$ increase compared to the heuristic and DRL schemes, which demonstrates the effectiveness of the proposed scheme. 

Fig.~\subref*{11b} presents the QoE variation with the increased bandwidth. {The proposed scheme's upward tendency in QoE with increased bandwidth implies that it can well adapt to varying network capacities}, which is essential for ensuring service quality during peak usage times or in bandwidth-constrained environments. When the bandwidth reaches $14~\myunit{MHz}$, the proposed scheme achieves a QoE increment of $11.1\%$ compared to the DRL scheme, suggesting that the proposed scheme can efficiently utilize available bandwidth to enhance the user experience. 

Fig.~\subref*{11c} correlates the computing capacity with QoE. {In MSVS, the computing capacity directly influences the transcoding version and the speed for segments to be delivered to \acp{smg}.} {The proposed scheme's pronounced improvement in QoE with the increasing computing capacity reflects its ability to leverage additional computational resources effectively for video transcoding management.} Here, the proposed scheme reached a QoE of around $1.75$ with a computing capacity of $12~\myunit{Gcycles/s}$, which is a significant $16.2\%$ improvement over the DRL method. {This suggests that the proposed scheme can effectively harness computing power to enhance video quality for QoE improvement.}

\begin{figure*}[!t]
\begin{align}
\label{yg}
\widehat{\Upsilon }_{g}^{{}}\left( {{a}_{g}}^{*},{{\beta }_{g}} \right)&={{\omega }_{g}}\left( \sum\limits_{m\in {{\Omega }_{g}}}{-1+\frac{1}{2a_{g,m}^{{{l}^{*}}}\sum\nolimits_{j=1}^{{{l}^{*}}}{z_{g,m}^{j}/\tau}+1}}+\frac{\lambda _{g,2}}{\left| {{\Omega }_{g}} \right|}\sum\limits_{m\in {{\Omega }_{g}}}{\left| Q_{g,m}^{{{l}^{*}}}\left( a_{g,m}^{{{l}^{*}}} \right)-Q_{g,m-1}^{{{l}^{*}}}\left( a_{g,m-1}^{{{l}^{*}}} \right) \right|} \right) \notag \\ 
&\qquad \qquad +{{\omega }_{g}}\lambda _{g,1}{{\left[ \max \left\{ \frac{\sum\nolimits_{m\in {{\Omega }_{g}}}{a_{g,m}^{{{l}^{*}}}\sum\nolimits_{j=1}^{{{l}^{*}}}{z_{g,m}^{j}}}}{\underset{k\in \bigcup\nolimits_{d=1}^{g}{{{\mathcal{K}}_{d}}}}{\mathop{\min }}\,{{\beta }_{g}}B{{\log }_{2}}\left( 1+\frac{{{\left| h_{g,k}^{{}} \right|}^{2}}{{P}_\text{D}}}{{{N}_{0}}} \right)},\frac{\mu \sum\nolimits_{m\in {{\Omega }_{g}}}{a_{g,m}^{{{l}^{*}}}\sum\nolimits_{j=1}^{{{l}^{*}}}{z_{g,m}^{j}}}}{{{\beta }_{g}}C} \right\}-\widetilde{q}_{g}^{0} \right]}^{+}}.
\end{align}
\noindent\rule{\textwidth}{0.4pt}  
\end{figure*}

Furthermore, we present the QoE comparison under different sensitivity degrees among different schemes in Fig.~\ref{fig:lambda}. As shown in Fig.~\subref*{12a}, we adjust the \acp{smg}'s sensitivity degrees of video rebuffering time with the fixed sensitivity degrees of video quality variation. The parameter $\lambda_1^1$ includes three elements, i.e., $(0.4, 0.3, 0.2)$, which represents three \acp{smg}' sensitivity degrees of rebuffering time, respectively. The parameter $\lambda_2^1$ also includes three elements, i.e., $(0.7, 0.6, 0.5)$. While the parameters $\lambda_1^2$ and $\lambda_1^3$ are set to $(0.3, 0.3, 0.3)$ and $(0.2, 0.3, 0.4)$, respectively. Different settings of sensitivity degrees of rebuffering time aim at validating the effectiveness of the proposed scheme in the diversified \acp{smg}. It can be observed that our proposed scheme can always achieve the highest QoE with a comparatively tighter range of variance under different parameters $\bm{\lambda_1}$. As shown in Fig.~\subref*{12b}, we change the \acp{smg}' sensitivity degrees of video quality variation with the fixed sensitivity degrees of rebuffering time. The parameters $\lambda_2^2$ and $\lambda_2^3$ are set to $(0.6, 0.6, 0.6)$ and $(0.5, 0.6, 0.7)$, respectively. Compared with the first error bar, the last two error bars obtain higher QoE values with similar variance. This is because the last two \acp{smg} are much more sensitive to video quality variation and they need to be allocated more bandwidth and computing resources to guarantee their smooth and high-quality video playback. Therefore, with a higher weighting factor, \acp{smg}' QoE can be effectively enhanced.

\section{Conclusion}
\label{sec:Conclusion}

In this paper, we have proposed a novel DT-based network management scheme to enhance QoE in MSVS. Furthermore, we have established a multicast QoE model to quantify the impact of multicast segment buffering among SMGs. A convex optimization embedded DRL algorithm has been designed to determine the joint segment version selection and slot division. The proposed DT-based network management scheme can efficiently multicast segment buffering in MSVS. For the future work, we will investigate the adaptive granularity of DT data collection and abstraction to reduce the network overhead of DT.

\appendices

\section{Proof of Theorem~\ref{theo1}}
\label{app:theo1}
When the variable $\bm{a}$ is determined, the opposite objective function in the minimization problem can be expressed in Eq.~\eqref{yg}.

Let denote 
\begin{equation}
    {{\Xi }_{1}}\left( {{\beta }_{g}} \right)=\left\{ \frac{\sum\nolimits_{m\in {{\Omega }_{g}}}{a_{g,m}^{{{l}^{*}}}\sum\nolimits_{j=1}^{{{l}^{*}}}{z_{g,m}^{j}}}}{\underset{k\in \bigcup\nolimits_{d=1}^{g}{{{\mathcal{K}}_{d}}}}{\mathop{\min }}\,{{\beta }_{g}}B{{\log }_{2}}\left( 1+\frac{{{\left| h_{g,k}^{{}} \right|}^{2}}{{P}_{D}}}{{{N}_{0}}} \right)} \right\} \notag
\end{equation} and 
\begin{equation}
    {{\Xi }_{2}}\left( {{\beta }_{g}} \right)=\left\{ \frac{\mu \sum\nolimits_{m\in {{\Omega }_{g}}}{a_{g,m}^{{{l}^{*}}}\sum\nolimits_{j=1}^{{{l}^{*}}}{z_{g,m}^{j}}}}{{{\beta }_{g}}C} \right\}, \notag
\end{equation} then the last term of Eq.~\eqref{yg} can be transformed into 
\begin{equation}
    {{\omega }_{g}}\lambda _{g,1}\left( \max \left\{ {{\left[ {{\Xi }_{1}}\left( {{\beta }_{g}} \right)-\widetilde{q}_{g}^{0} \right]}^{+}},{{\left[ {{\Xi }_{2}}\left( {{\beta }_{g}} \right)-\widetilde{q}_{g}^{0} \right]}^{+}} \right\} \right).\notag
\end{equation}

Since the second derivative of functions $\Xi_1(\beta_g)$ and $\Xi_2(\beta_g)$ are positive values, i.e., 
$\frac{{{\partial }^{2}}{{\Xi }_{1}}\left( {{\beta }_{g}} \right)}{{{\partial }^{2}}{{\beta }_{g}}}\ge 0,\frac{{{\partial }^{2}}{{\Xi }_{2}}\left( {{\beta }_{g}} \right)}{{{\partial }^{2}}{{\beta }_{g}}}\ge 0$, they are convex functions. Then, we need to prove the convexity of function ${{\psi }_{1}}({{\beta }_{g}})={{\left[ {{\Xi }_{1}}\left( {{\beta }_{g}} \right)-\widetilde{q}_{g}^{0} \right]}^{+}}$ and ${{\psi }_{2}}({{\beta }_{g}})={{\left[ {{\Xi }_{2}}\left( {{\beta }_{g}} \right)-\widetilde{q}_{g}^{0} \right]}^{+}}$.

Consider an arbitrary value $\theta \in (0,1)$, and arbitrary values $\beta _{g,1}$, and $\beta _{g,2}$, we have
\begin{align}
&{{\psi }_{1}}(\theta \beta _{g,1}+(1-\theta )\beta _{g,2})\notag \\
&\quad =\max \{{{\Xi }_{1}}\left( \theta \beta _{g,1}+(1-\theta )\beta _{g,2} \right)-\widetilde{q}_{g}^{0},0\}\notag \\ 
&\quad\le \max \{\theta {{\Xi }_{1}}(\beta _{g,1})+(1-\theta ){{\Xi }_{1}}(\beta _{g,2})-\widetilde{q}_{g}^{0},0\}\notag \\ 
&\quad\le \theta \max \{{{\Xi }_{1}}(\beta _{g,1})-\widetilde{q}_{g}^{0},0\} \notag \\
&\phantom{\theta \max \{{{\Xi }_{1}}(\beta _{g,1})-\widetilde{q}_{g}^{0},0\}} +(1-\theta )\max \{{{\Xi }_{1}}(\beta _{g,2})-\widetilde{q}_{g}^{0},0\}\notag \\ 
&\quad=\theta {{\psi }_{1}}(\beta _{g,1})+(1-\theta ){{\psi }_{1}}(\beta _{g,2}).
\end{align}

Therefore, the function ${{\psi }_{1}}({{\beta }_{g}})$ is convex. The same validation method can be applied to ${{\psi }_{2}}({{\beta }_{g}})$ and $\max \left\{ {{\psi }_{1}}({{\beta }_{g}}),{{\psi }_{2}}({{\beta }_{g}}) \right\}$. Therefore, the component ${{\omega }_{g}}\lambda _{g,1}\left( \max \left\{ {{\psi }_{1}}({{\beta }_{g}}),{{\psi }_{2}}({{\beta }_{g}}) \right\} \right)$ is convex. Since the first two terms of the first line of Eq.~\eqref{yg} are constant, they do not affect the convexity of the transformed objective function. Therefore, function $\widehat{\Upsilon }_{g}^{{}}\left( {{a}_{g}}^{*},{{\beta }_{g}} \right)$ is convex. For functions $\widehat{\Upsilon }_{g}^{{}}\left( {{a}_{g}}^{*},{{\beta }_{g}} \right)$ with different $g$, they have an identical domain and are mutually independent with each other, so their summation is also a convex function. Based on the above analysis, the transformed objective function is convex.

%
\bibliographystyle{IEEEtran}
\bibliography{IEEEabrv,Ref}
%
%

\end{document}

%% file: Abbreviation_Definitions.tex
\DeclareAcronym{SNR}{
	short = SNR,
	long = signal-to-noise ratio}
 \DeclareAcronym{ap}{
	short = AP,
	long = access point}
 \DeclareAcronym{udt}{
	short = UDT,
	long = user digital twin}
 \DeclareAcronym{mg}{
	short = MG,
	long = multicast group}
  \DeclareAcronym{smg}{
	short = SMG,
	long = sub-MG}
   \DeclareAcronym{msvs}{
	short = MSVS,
	long = multicast short video streaming}
    \DeclareAcronym{svc}{
	short = SVC,
	long = scalable video coding}
\DeclareAcronym{ssim}{
	short = SSIM,
	long = structural similarity index measure}
\DeclareAcronym{drl}{
	short = DRL,
	long = deep reinforcement learning}
\DeclareAcronym{bdq}{
	short = BDQN,
	long =  branching dueling Q network}
\DeclareAcronym{slsqp}{
	short = SLSQP,
	long = sequential least squares quadratic programming}
 \DeclareAcronym{QoE}{
	short = QoE,
	long = quality of experience}
 \DeclareAcronym{dt}{
	short = DT,
	long = digital twin}
 \DeclareAcronym{hdcp}{
	short = HDCP,
	long = hard deadline constrained prioritized}
 \DeclareAcronym{sinr}{
	short = SINR,
	long = signal to interference plus noise ratio}